\documentclass[12pt]{article}
\usepackage{subcaption}
\usepackage{graphicx}
\usepackage{amsmath, nccmath}
\usepackage{amsfonts}
\usepackage{amssymb}
\usepackage{amsthm}
\usepackage{mathdots}
\usepackage{color}
\usepackage{graphicx}
\usepackage{tikz}
\usepackage{dsfont}
\usepackage{mathbbol}
\usepackage{amssymb} 
\usepackage{simpler-wick}
\DeclareSymbolFontAlphabet{\amsmathbb}{AMSb}%
\usetikzlibrary{shapes.misc}
\usepackage{jheppub} 
\usepackage{physics}
\usepackage{mathtools}
\usepackage{tensor}
\usepackage{mathrsfs}
\usepackage{comment}
\usepackage{bbm}
\usepackage{graphicx}
\usepackage{ragged2e}
\usepackage{color}
\usepackage{wrapfig}

\tikzset{cross/.style={cross out, draw=black, minimum size=2*(#1-\pgflinewidth), inner sep=0pt, outer sep=0pt},
cross/.default={3pt}}
\usetikzlibrary{arrows,shapes,positioning}
\usetikzlibrary{decorations.markings}
\usepackage[rightcaption]{sidecap}
\tikzstyle arrowstyle=[scale=1]
\tikzstyle directed=[postaction={decorate,decoration={markings,
    mark=at position .65 with {\arrow[arrowstyle]{stealth}}}}]
\tikzstyle reverse directed=[postaction={decorate,decoration={markings,
    mark=at position .65 with {\arrowreversed[arrowstyle]{stealth};}}}]
\usetikzlibrary{positioning}

\renewcommand{\thanks}[1]{\footnote{#1}}

\newcommand{\bea}{\begin{eqnarray}}
\newcommand{\eea}{\end{eqnarray}}
\newcommand{\be}{\begin{eqnarray}}
\newcommand{\ee}{\end{eqnarray}}
\newcommand{\bma}{\begin{matrix}}
\newcommand{\ema}{\cr\end{matrix}}

\newcommand{\SL}{ {\text {SL}}}

\newcommand{\SO}{ {\text {SO}}}
\newcommand{\U}{ {\text {U}}}
\newcommand{\tO}{ {\text {O}}}

\def\normOrd#1{\mathop{:}\nolimits\!#1\!\mathop{:}\nolimits}
\newcommand{\reals}{ {\amsmathbb R} }
\newcommand{\naturals}{ {\amsmathbb N} }
\newcommand{\complex}{ {\amsmathbb C} }
\newcommand{\integers}{ {\amsmathbb Z} }
\newcommand{\rationals}{ {\amsmathbb Q} }

\def\cA{{\cal A}}
\def\cB{{\cal B}}

\def\cD{{\cal D}}

\def\cF{{\cal F}}
\def\cG{{\cal G}}

\def\cM{{\cal M}}

\def\cO{{\cal O}}
\def\cP{{\cal P}}

\def\cR{{\cal R}}

\def\cT{{\cal T}}

\def\cZ{{\cal Z}}

\def\Im{{\rm Im \,}}

\def\det{{\rm det \,}}

\def\half{{1\over 2}}

\def\a{\alpha}
\def\b{\beta}
\def\g{\gamma}
\def\d{\delta}

\def\g{\gamma}

\def\ep{\varepsilon}

\def\no{\nonumber}

\definecolor{darkred}{rgb}{0.8,0.1,0.1}
\hypersetup{colorlinks=true, linkcolor=darkred, citecolor=blue, linktoc=page}

\newcommand{\nc}{\newcommand}
\nc{\rnc}{\renewcommand} 

\rnc{\a}{\alpha}
\rnc{\b}{\beta}
\rnc{\d}{\delta}
\nc{\e}{\epsilon}
\nc{\z}{\zeta}
\nc{\m}{\mu}
\nc{\n}{\nu}
\rnc{\r}{\rho}
\rnc{\k}{\kappa}
\rnc{\l}{\lambda}
\nc{\s}{\sigma}
\rnc{\t}{\tau}
\nc{\w}{\omega}
\nc{\x}{\chi}
\nc{\F}{\Phi}
\rnc{\L}{\Lambda}

\nc{\pd}{\partial}

\usepackage[draft,inline,nomargin,index]{fixme}
\usepackage{comment}
\FXRegisterAuthor{kr}{akr}{\color{blue}KR}

\title{Non-invertible defects on the worldsheet}

\author[a]{Sriram Bharadwaj,}
\author[a]{Pierluigi Niro,}
\author[b]{and Konstantinos Roumpedakis}

\affiliation[a]{
Mani L. Bhaumik Institute for Theoretical Physics, Department of Physics and Astronomy,\\ University of California, Los Angeles, CA 90095, U.S.A.
}
\affiliation[b]{William H. Miller III Department of Physics and Astronomy, Johns Hopkins University,\\ 3400 North
Charles Street, Baltimore, MD 21218, U.S.A.
}

\emailAdd{sbharadwaj@physics.ucla.edu}
\emailAdd{pniro@physics.ucla.edu}
\emailAdd{kroumpe1@jh.edu}

\abstract{
We consider codimension-one defects in the theory of $d$ compact scalars on a two-dimensional worldsheet, acting linearly by mixing the scalars and their duals.
By requiring that the defects are topological, we find that they correspond to a non-Abelian zero-form symmetry acting on the fields as elements of $\tO(d;\reals) \times \tO(d;\reals)$, and on momentum and winding charges as elements of $\tO(d,d;\reals)$.
When the latter action is rational, we prove that it can be realized by combining gauging of non-anomalous discrete subgroups of the momentum and winding $\U(1)$ symmetries, and elements of the $\tO(d,d;\integers)$ duality group, such that the couplings of the theory are left invariant.
Generically, these defects map local operators into non-genuine operators attached to lines, thus corresponding to a non-invertible symmetry.
We confirm our results within a Lagrangian description of the non-invertible topological defects associated to the $\tO(d,d;\rationals)$ action on charges, giving a natural explanation of the rationality conditions.
Finally, we apply our findings to toroidal compactifications of bosonic string theory. In the simplest non-trivial case, we discuss the selection rules of these non-invertible symmetries, verifying explicitly that they are satisfied on a worldsheet of higher genus.
}

\begin{document}
\maketitle

\newpage
\section{Introduction}
In recent years, generalized symmetries in various dimensions have been a very active field of investigation \cite{Gaiotto:2014kfa,Bhardwaj:2017xup,Chang:2018iay,Lin:2019hks,Thorngren:2019iar,Komargodski:2020mxz, Thorngren:2021yso, Nguyen:2021yld, Choi:2021kmx, Kaidi:2021xfk, Choi:2022zal, Cordova:2022ieu, Roumpedakis:2022aik,Bhardwaj:2022yxj,Damia:2022rxw,Arias-Tamargo:2022nlf,Hayashi:2022fkw,Kaidi:2022uux,Antinucci:2022eat, Choi:2022rfe, Choi:2022jqy,Bartsch:2022mpm, Heidenreich:2021xpr, Cordova:2022rer,Huang:2021zvu, Bashmakov:2022jtl,Benini:2022hzx, Damia:2022bcd,Lin:2022dhv, Niro:2022ctq,Cordova:2023ent,Sela:2024okz, Rudelius:2020orz, Inamura:2021wuo, Nguyen:2021naa, Bhardwaj:2022lsg, Kaidi:2022cpf, Chen:2022cyw, Bashmakov:2022uek, Cordova:2022fhg, Choi:2022fgx,Yokokura:2022alv,Bhardwaj:2022kot,Bhardwaj:2022maz,Bartsch:2022ytj,Apte:2022xtu,Kaidi:2023maf,Koide:2023rqd,Damia:2023ses,Bhardwaj:2023ayw,vanBeest:2023dbu,Lawrie:2023tdz,Chen:2023czk,Cordova:2023bja,Antinucci:2023ezl,Benedetti:2023owa,Choi:2023pdp,Nagoya:2023zky,Okada:2024qmk,Kan:2024fuu,Gutperle:2024vyp} providing valuable insights into the dynamics of quantum field theories. In \cite{Gaiotto:2014kfa}, a definition of symmetry was given in terms of topological operators. This point of view has been proved to be quite useful for exploring more exotic types of symmetries, which are generated by topological operators that do not necessarily have an inverse, and thus are referred to as non-invertible symmetries.
The goal of this paper is to explore non-invertible symmetries on the worldsheet in string theory, focusing on the two-dimensional theory of $d$ compact scalars with a non-trivial metric and a $B$-field, which indeed arises from a toroidal compactification of bosonic string theory.

In two-dimensional conformal field theories (CFTs), non-invertible symmetries have a long history, starting with the work of Verlinde in rational CFTs \cite{Verlinde:1988sn}. For theories with $c\geq 1$, Verlinde lines are defined as topological defects which preserve an affine algebra. Even in such theories, there exist more general topological defects, which only commute with the Virasoro algebra and not with the whole current algebra. 

RCFTs can be alternatively defined by specifying the fusion category of topological defects \cite{Fuchs:2002cm}. In this way, the topological defects are part of the definition of the theory. For theories defined by a Lagrangian, although we do not yet have a general way to look for non-invertible symmetries, there are a few methods that apply in special cases. If the theory possesses duality transformations, a way to construct topological defects is by the half-space gauging approach of \cite{Choi:2021kmx}. In this approach, we can construct topological defects by combining duality transformations and gauging of non-anomalous discrete symmetries in half-space, such that the couplings of the theory are left invariant. A natural setup to apply these methods is the worldsheet theory in string compactifications, which exhibits a large set of dualities depending on the choice of the target space. For related past work, see \cite{Frohlich:2006ch, Fuchs:2007tx, Bachas:2012bj, Thorngren:2021yso, Nagoya:2023zky, Damia:2024xju}. 

It is well-known that the toroidal compactification of bosonic string theory enjoys an $\tO(d,d;\mathbbm Z)$ duality group (for a review see e.g.~\cite{Giveon:1994fu}). The goal of this work is to show how duality elements can be combined with gaugings of discrete subgroups of the momentum and winding symmetries to construct topological defects. We show that when the target space metric and the $B$-field obey certain rationality conditions, we can construct an infinite number of topological defects corresponding to elements of $\tO(d) \times \tO(d)$ acting on the scalars. These act on momentum and winding charges as elements of $\tO(d,d;\rationals)$ and have, generically, a non-trivial kernel, and thus correspond to non-invertible symmetries. Our work generalizes to arbitrary $d$ the fact that the theory of a single compact boson with radius $R$ enjoys a non-invertible symmetry at rational values of $R^2=p/q$, obtained by combining T-duality and the gauging of a $\integers_p \times \integers_q$ subgroup of the zero-form symmetry. This defect acts as $\tO(1)\times\tO(1)\cong\integers_2 \times \integers_2$ on $\partial\phi$ and $\Bar{\partial}\phi$, and as an element of $\tO(1,1;\rationals)$ on the charges.

On the string worldsheet, topological defects act on massless states and therefore flow to non-trivial defects in the low-energy effective theory. For recent work on topological defects on the worldsheet, see \cite{Cordova:2023qei,Kaidi:2024wio, Heckman:2024obe}. A feature of non-invertible defects is that they lead to different selection rules on worldsheets of different topology. However, the modified selection rules are satisfied order by order in string perturbation theory, whereas in the full non-perturbative setup one does not expect global symmetries to be present in a theory of quantum gravity. 

This work is organized as follows. In section \ref{section: Topological defects on the worldsheet}, we start with a brief review and then we present the main result of this work. We focus on string theory compactified on a $d$-dimensional torus, and we consider the most general (linear) transformations of the target space coordinates that can arise from codimension-one defects. By imposing that the defects are topological, we find that they act on the target space coordinates as elements of $\tO(d)\times\tO(d)$ and on the global charges as elements of $\tO(d,d)$. Then, we prove that the transformations corresponding to an $\tO(d,d;\rationals)$ action on charges can be realized in the half-space gauging approach. In turn, this guarantees the existence of the corresponding topological defects in the quantum theory. 
We also discuss the fusion rules of these defects, showing that they are in general non-invertible as well as non-Abelian. 

In section \ref{section: Lagrangian description of the topological defects}, we take an alternative approach and we present a Lagrangian description of the topological defects we constructed. We find that the rationality constraints of the previous section are required by gauge invariance of the defect Lagrangian. We then work out an example of a fusion between two defects and verify the results from the general approach of the previous section.  

In section \ref{section: defects at d=1}, we explore the constraints of these topological defects in string theory. For simplicity, we focus on bosonic string theory with only one compactified direction.
We argue that the topological defects on the worldsheet lead to global symmetries in the low-energy target space effective theory, whose selection rules follow from string amplitudes on the sphere. In string perturbation theory, the selection rules of non-invertible topological defects on higher-genus worldsheets are different with respect to the ones on the sphere. We consider a string correlator on the torus and prove explicitly that the modified selection rules are satisfied in an example of a non-trivial one-point function. Finally, we comment about the role of these worldsheet non-invertible defects in the context of string perturbation theory.
\section{Topological defects on the worldsheet}
\label{section: Topological defects on the worldsheet}
\subsection{Generalities}

Consider a two-dimensional bosonic worldsheet theory defined on a target space with a metric $G_{IJ}$, an antisymmetric $B$-field $B_{IJ}$, and a dilaton $\Phi_D$. In the conventions outlined in appendix \ref{conventions}, the action (with $\a'=1$) is given by
\begin{align}
    S_D[X]=\frac{1}{4\pi} \int_\Sigma \left( G_{IJ}\,dX^I\wedge\star dX^J + iB_{IJ}\,dX^I\wedge dX^J - \frac{1}{2}\mathcal{R}_g \Phi_D \, \text{Vol}_\Sigma \right) ~,
\end{align}
where $\Sigma$ is the 2d worldsheet with Ricci scalar $\mathcal{R}_g$ and volume form $\text{Vol}_\Sigma$, parametrized by Euclidean coordinates $(x,y)$. The indices $I, J =0,1,\dots, D-1$ label the target space coordinates $X^I$. In matrix notation, the holomorphic stress tensor takes the form 
\begin{align}
    T_D = -\frac{1}{2}(\partial X)^T G(\partial X) ~, \qquad \Bar{T}_D = -\frac{1}{2}(\Bar\partial X)^T G(\Bar\partial X) ~,
\end{align}
where $\partial$ ($\Bar\partial$) denotes the derivative with respect to the complex coordinate $z=x+iy$ ($\Bar{z}=x-iy$). Suppose that we compactify $d$ dimensions of the target space on a $d$-dimensional torus $\amsmathbb{T}^d$. We denote the compact directions as $\phi^k$, with $k = 1,\dots, d$. We take all scalars to be $2\pi$ periodic, namely $\phi^k \sim \phi^k+2\pi$, and we absorb the geometrical data of the torus in $G$ and $B$. Furthermore, we assume that the metric and the $B$-field split as 
\begin{align}
    G_{IJ}\rightarrow \text{diag}(G_{\mu\nu}, G_{ij})\,,  \quad B_{IJ}\rightarrow \text{diag}(B_{\mu\nu}\,, B_{ij})~, \quad
    \Phi_D \rightarrow \Phi_d + \frac{1}{2}\log\det G_{ij} \,,
\end{align}
where $\mu,\nu=0,1,\dots, D-1-d$, and $G_{ij},B_{ij}\in\reals$ are constant symmetric and antisymmetric matrices, respectively, with $i,j=1,\dots,d$. This splitting of the target space metric and the $B$-field implies that compact and non-compact scalars decouple.

Henceforth, we focus on the compact sector of the theory, whose action and holomorphic stress tensor are written in terms of the $d$ compact scalars $\phi^i$ as\footnote{For our discussion, we can safely neglect the contribution of the dilaton to the action of the compact sector. Indeed, this gives a contribution proportional to the Euler characteristic $\chi_g$ of $\Sigma$,
\begin{equation*}
S_\Phi = - \frac{1}{16\pi} \log\det G_{ij}  \int_\Sigma \mathcal{R}_g \text{Vol}_\Sigma = -\frac{\chi_g}{4} \log\det G_{ij} \,. 
\end{equation*}
As it does not depends on $\phi^i$, topological symmetry defects (which preserve the couplings $G$ and $B$) clearly leave this term invariant.
}
\begin{align}
    S[\phi]=\frac{1}{4\pi} \int_\Sigma \left( G_{ij}\, d\phi^i\wedge\star d\phi^j + i B_{ij}\, d\phi^i\wedge d\phi^j \right) \,, \qquad T =-\frac{1}{2}(\partial\phi)^TG(\partial\phi)~.
    \label{eq:compact_action}
\end{align}
The theory \eqref{eq:compact_action} enjoys several zero-form symmetries. For each compact direction, there is a $\U(1)_m$ momentum symmetry that shifts the corresponding scalar by a constant
\begin{align}
    \U(1)_{m}^{(i)}:\phi^i\rightarrow\phi^i+ c^i~.
\end{align}
The Noether currents and the associated charges are
\begin{align}
    (\star j_m)_i & = \frac{i}{2\pi} G_{ij}\star d\phi^j -\frac{1}{2\pi} B_{ij}d\phi^j \equiv \frac{1}{2\pi}d\Tilde\phi_i\,, & m_i&=\int_\gamma (\star j_m)_i~,
    \label{eq:momentumJ}
\end{align}
where we have also defined the dual scalars $\Tilde\phi_i$, which are compact with period $2\pi$.
Similarly, for each compact direction there is a winding $\U(1)_w$ symmetry associated to shifts of the dual scalars
\begin{align}
    \U(1)_{w}^{(i)}:\Tilde\phi_i\rightarrow\Tilde\phi_i+ \Tilde{c}_i~.
\end{align}
The winding currents and charges are
\begin{align}
    (\star j_w)^i &= \frac{1}{2\pi}d\phi^i~,& w^i&=\int_\gamma (\star j_w)^i~.
    \label{eq:windingJ}
\end{align}
In summary, the continuous part of the symmetry group is
\begin{align} \label{eq:symmetries}
\prod_{i=1}^d \U(1)_{m}^{(i)}\times\U(1)_{w}^{(i)}~,
\end{align}
associated to a vector $\mathbf{q}\equiv (w^1,\dots, w^d,m_1,\dots,m_d)^T$ of $2d$ integer charges.
Furthermore, there is also a discrete reflection symmetry, acting as $\phi^i \rightarrow -\phi^i$ simultaneously on all scalars.

\paragraph{Dualities.} It is well-known that the theory \eqref{eq:compact_action} has an $\text{O}(d,d;\integers)$ duality group, see e.g.~\cite{Giveon:1994fu}. In general, a duality transformation is \textit{not} a symmetry since it acts on the couplings. The duality group $\text{O}(d,d;\integers)$ can be conveniently represented as the group of $2d\times 2d$ matrices $\cM$ with integer entries such that 
\begin{align}
    \cM^T J \cM = J~, \qquad \text{ with } J = \begin{pmatrix}
        0&\mathbb{1}_d\\
        \mathbb{1}_d&0
    \end{pmatrix}~.
\end{align}
The action of an element $\cM$ of the duality group on the fields $\mathbf{d\Phi}=(d\phi^i,d\Tilde\phi_j)^T$ and on the vector of charges $\mathbf{q}=(w^i,m_j)^T$ is
\begin{align}\label{OddOnq}
\mathbf{d\Phi}\rightarrow \cM^{-T} \, \mathbf{d\Phi}~,\qquad    \mathbf{q}\rightarrow \cM^{-T} \, \mathbf{q}~,\qquad \cM^{-T}=J\cM J \,,
\end{align}
which corresponds to the following action on the couplings (encoded in the matrices $G$ and $B$)
\begin{align}\label{OddOnZ}
    Z(G,B) \rightarrow \cM \, Z(G,B) \, \cM^T ~.
\end{align}
The generalized metric $Z(G,B)\in \text{O}(d,d;\reals)$ is the symmetric matrix given by \cite{Giveon:1994fu}
\begin{align}\label{couplings}
    Z(G,B) =
    \begin{pmatrix}
        \mathbb{1}_d&\,-B\\
        0&\mathbb{1}_d
    \end{pmatrix}
    \begin{pmatrix}
        G&0\\
        0&G^{-1}
    \end{pmatrix}
    \begin{pmatrix}
        \mathbb{1}_d&0\\
        B&\mathbb{1}_d
    \end{pmatrix}
    =
    \begin{pmatrix}
        G - BG^{-1}B\;\; &-BG^{-1}\\
        G^{-1}B\;\;&G^{-1}
    \end{pmatrix}~.
\end{align}
The theory is invariant by combining the two operations \eqref{OddOnq} and \eqref{OddOnZ}, which constitute the duality transformations acting both on the charges and on the couplings. This can be easily checked since the zero-mode Hamiltonian is $H=\frac{1}{2}\mathbf{q}^T Z(G,B) \mathbf{q}$.

Let us review the generating elements of the duality group $\text{O}(d,d;\integers)$. One can write any element of the duality group as a finite product of generating elements. They are organized in three types, corresponding to the action on $G$ and $B$ described below.
\begin{enumerate}
    \item Integer shift of $B_{ij}$ by an antisymmetric integer matrix $\Theta_{ij}$. This is implemented by a transformation of the form
    \begin{align}
        \cM_\Theta &=\begin{pmatrix}
            \mathbb{1}_d & -\Theta\\
            0 &\mathbb{1}_d
        \end{pmatrix}~.
    \label{Bshiftmatrix}    
    \end{align}
    \item Basis change of the compactification lattice by an integer matrix $A\in\text{GL}(d,\integers)$, corresponding to $E\rightarrow AEA^T$, where $E = G+B$. This is implemented by
    \begin{align}
        \cM_A =\begin{pmatrix}
            A&0\\
            0&A^{-T}
        \end{pmatrix}~.
    \label{basischangematrix}    
    \end{align}
    \item Generalized T-dualities, implemented by the matrices
    \begin{align}
        \cM_i=\begin{pmatrix}
            \mathbb{1}_d-e_i& e_i\\
            e_i&\mathbb{1}_d-e_i
        \end{pmatrix} \,,\quad \text{ where }(e_i)_{k\ell} = \delta_{k,i}\delta_{\ell,i}\,, \quad i,k,\ell=1,\dots,d~.
    \label{generalizedTmatrix}   
    \end{align}
    When $G=\text{diag}(R_1^2,\dots,R_d^2)$ and $B=0$, this reduces to the usual T-duality acting on the single $i^{\rm th}$ scalar as $\phi^i \leftrightarrow \Tilde{\phi}_i$ and on its radius as $R_i \rightarrow 1/R_i$ .    
\end{enumerate}

\paragraph{Gauging.} For each $i=1,\dots,d$, the corresponding $\U(1)_m^{(i)}\times\U(1)_w^{(i)}$ symmetry has a mixed 't Hooft anomaly, which can be seen by coupling the theory to background gauge fields for both symmetries. Any $\integers_{p_i}\times\integers_{q_i}$ subgroup of $\U(1)_m^{(i)}\times \U(1)_w^{(i)}$ with $p_i,q_i\in\naturals$ and $\gcd(p_i,q_i)=1$ is anomaly-free and can be gauged.
This gauging acts on the vector $\mathbf{q}$ of integer charges as
\begin{align}\label{gauging}
    \mathbf{q}\rightarrow \begin{pmatrix}
        \cG^{-1}&0\\
        0&\cG
    \end{pmatrix} \mathbf{q} \,, \qquad \text{ with } \cG& = \text{diag} \left(\frac{q_1}{p_1},\frac{q_2}{p_2},\dots,\frac{q_d}{p_d}\right).
\end{align}
Notice that each diagonal entry of $\cG$ is a rational number, so that not every integer $\mathbf{q}$ is mapped to an integer $\mathbf{q}$. This implies that gauging projects out states whose charges are not integers after gauging. This property is crucial to characterize the non-invertible nature of the symmetry defects that can be constructed by combining gauging and dualities, as we explain in the next section.

The gauging matrix acting on the charges as in \eqref{gauging} is a matrix of $\text{O}(d,d;\rationals)$, since
\begin{align}\label{GisOdd}
    \begin{pmatrix}
         \mathcal G^{-1} & 0\\
        0& \mathcal G
    \end{pmatrix}
    J
    \begin{pmatrix}
         \mathcal G^{-1} & 0\\
        0& \mathcal G
    \end{pmatrix} = J ~.
\end{align}
This implies that gauging operations, as well as duality transformations, preserve the Dirac quantization condition
\be
q^TJq \in 2\integers \,,
\ee
as this combination is precisely the invariant scalar product of $\tO(d,d)$.

Gauging a discrete subgroup of the global symmetry can be either viewed as a transformation which acts on the periodicities of the compact scalars, leaving the couplings invariant, or as a transformation which acts on the couplings but does not change the periodicities.\footnote{This is analogous to what happens in the familiar case of 4d $\U(1)$ Maxwell theory with complexified gauge coupling $\tau$. For instance, gauging a $\integers_N$ subgroup of the electric one-form symmetry maps the theory to a $\U(1)/\integers_N$ Mawwell theory with the same coupling $\tau$. After a field redefinition, this is equivalent to a $\U(1)$ Maxwell theory with coupling $\tau/N^2$. The advantage of this presentation is that gauging can be viewed as a map acting on the coupling (and on the charges), without changing the periodicity of the gauge field, i.e.~the gauge group.} We adopt the latter viewpoint here.
In particular, the action of the gauging operation \eqref{gauging} on the couplings is $G\rightarrow \cG G \cG $ and $B\rightarrow \cG B \cG$. Therefore, on $Z(G,B)$ it acts as 
\begin{align}\label{GonZ}
    Z(G,B) \rightarrow
    \begin{pmatrix}
        \cG & 0\\
        0& \cG^{-1}
    \end{pmatrix}
    Z(G,B)
    \begin{pmatrix}
        \cG & 0\\
        0& \cG^{-1}
    \end{pmatrix}~.
\end{align}
\subsection{Topological defects from the stress tensor}
\label{section: topological defects from the stress tensor}

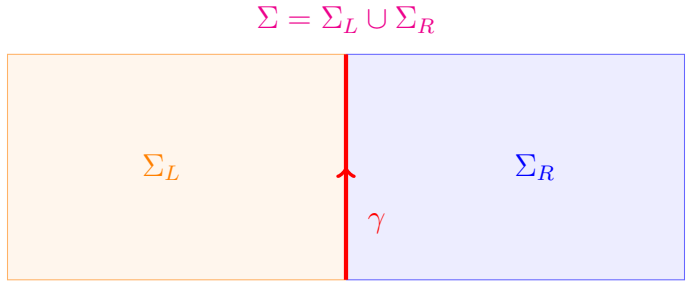
\begin{figure}[ht]
    \centering
    \begin{tikzpicture}[scale=1.5]
\filldraw[blue!60, fill=blue!7] (3,0) -- (6,0) -- (6,2) -- (3,2) ;
\draw[orange!60, fill=orange!7] (3,0) -- (0,0) -- (0,2) -- (3,2)  ;
\draw[red, ultra thick] (3,2)--(3,0);
\draw[red,->,ultra thick]        (3,1);
\node[label={0:\color{blue}$\Sigma_R$}] at (4.3,1) {};
\node[label={0:\color{orange}$\Sigma_L$}] at (1,1) {};
\node[label={0:\color{red}$\gamma$}] at (3,0.5) {};
\node[label={\color{magenta}$\Sigma = \Sigma_L \cup\Sigma_R$}] at (3,2) {};
\end{tikzpicture}
    \caption{A pictorial description of the worldsheet $\Sigma$ with a topological defect inserted along the line $\gamma$. For simplicity, we take $\gamma$ to be at $x=0$.}
    \label{WSpic}
\end{figure}

In this section, we study codimension-one topological defects in the theory \eqref{eq:compact_action}.
We consider a defect along a curve $\gamma$, which  splits the worldsheet as $\Sigma = \Sigma_L\cup\Sigma_R$, such that $\partial\Sigma_L = -\partial\Sigma_R = \gamma$. 
For definiteness, we consider, without loss of generality, a defect located at $x=0$, as depicted in figure \ref{WSpic}.
At the classical level, a defect is topological if the stress tensor is conserved across the defect itself, namely the holomorphic components need to satisfy
\begin{equation}\label{Tcons}
    T_L|_\gamma=T_R|_\gamma, \qquad \Bar{T}_L|_\gamma=\Bar{T}_R|_\gamma~,
\end{equation}
where $T_{L,R}$ denote the stress tensors which depend on the fields $\phi^i_{L,R}$ as in \eqref{eq:compact_action}, on the left and on the right of the defect, respectively.\footnote{With $\phi_{L,R}$ we refer to the fields on the left and on the right of the defect, respectively, and not to the left and right movers. The same applies to the vector of conserved charges $\mathbf{q}_{L,R}$. In our setup, the right fields/charges are acted upon by the defect to produce the left fields/charges.}
Our main goal in this section is to construct the most general solution to this pair of equations. Importantly, the couplings $G$ and $B$ are the same on the left and on the right of the defect.\footnote{If the couplings are not left invariant, the defect realizes an interface between two copies of the theory at different values of the couplings. Instead, here we are interested in topological defects realizing a symmetry operator in a given theory specified by $(G,B)$.} The latter acts on the fields via a generic linear transformation
\begin{align}
    \begin{pmatrix}
        \partial {\phi}_L|_\gamma\\
        \Bar{\partial} {\phi}_L|_\gamma
    \end{pmatrix} & = \begin{pmatrix}
        M&N_1\\
        N_2 &\Bar{M}
    \end{pmatrix}\begin{pmatrix}
        \partial \phi_R|_\gamma\\
        \Bar{\partial} \phi_R|_\gamma
    \end{pmatrix}~,
    \label{eq:action_phi}
\end{align}
where $\partial\phi$ and $\Bar\partial\phi$ are $d$-dimensional vectors, and $M$, $\Bar{M}$, $N_1$, and $N_2$ are independent $d\times d$ matrices with real coefficients.
Imposing the topological constraint (\ref{Tcons}) we get
\begin{align}\label{MGM}
    N_1=N_2=0 \,, \qquad M^T GM = G\,, \qquad \Bar{M}^TG\Bar{M}&=G~.
\end{align}
This means that a topological defect does not mix holomorphic with anti-holomorphic fields, and it rotates such components separately with a trasformation which is orthogonal with respect to the metric $G$. This shows that these defects correspond to $\tO(d;\reals)\times\tO(d;\reals)$ transformations.

From \eqref{eq:action_phi}, we can determine the action of the above classical topological defects on the conserved currents \eqref{eq:momentumJ} and \eqref{eq:windingJ}. For compactness, let us define on $\gamma$ the one-form 
\begin{equation}
    v_\complex \equiv \begin{pmatrix}
        \partial \phi|_\gamma\\
        \Bar{\partial}\phi|_\gamma
    \end{pmatrix}dy~.
\end{equation}
Using \eqref{MGM} we have that 
\begin{align}\label{eq:actiononv_C}
    v_\complex^L = M_\complex \, v_\complex^R\,, \qquad \text{ where } \;M_\complex & = \begin{pmatrix}
        M&0\\
        0 & \Bar{M}
    \end{pmatrix}~.
\end{align}
To determine the action on the conserved currents, we observe that they can be expressed in terms of $v_\complex$ as
\begin{align}\label{def_vq}
     v_q \equiv 2\pi\begin{pmatrix}
        \star j_w|_\gamma\\
        \star j_m|_\gamma
    \end{pmatrix}= U_B \, v_\complex~,
\end{align}
where\footnote{This relation follows from rewriting the conserved currents on the defects in complex coordinates, by using $d\phi|_\gamma=i(\partial\phi-\Bar\partial\phi)|_{x=0} \, dy$ and $i\star d\phi|_\gamma=i(\partial\phi+\Bar\partial\phi)|_{x=0} \, dy$.\label{basischange}}
\begin{align}
    U_B= i \begin{pmatrix}
        \mathbb{1}_d&-\mathbb{1}_d\\
        G-B\,\,&\,\, G+B
    \end{pmatrix}~.
\end{align}
We conclude that the conserved currents on the left and on the right of the defect are related by 
\begin{align}
    v_q^L = M_B \,v_q^R\,, \qquad  M_B \equiv U_B\,M_\complex\,U_B^{-1}~.
    \label{eq:action_on_Js}
\end{align}
Equivalently, one can integrate this equation to get the action of the defect on the vector of conserved charges as
\begin{equation}\label{eq:action_on_charges}
    \mathbf{q}_L = M_B \, \mathbf{q}_R~.
\end{equation}
By definition, $M_B$ acts on the charges but leaves the couplings invariant. It is straightforward to verify that $M_B$ preserves the generalized metric \eqref{couplings}, namely
\begin{equation}\label{eq:action_on_Z}
M_B^{T}\,Z(G,B)\,M_B = Z(G,B) \,,
\end{equation}
which implies that it leaves the zero-mode Hamiltonian $H=\frac{1}{2}\mathbf{q}^T Z(G,B) \mathbf{q}$ invariant, as expected for symmetry defects.

Furthermore, we have that $M_B \in \text{O}(d,d;\reals)$. To prove this, notice that the matrix $M_B$ can be rewritten as
\begin{align}\label{MBproduct}
    M_B = (U_0 U_B^{-1})^{-1}M_0(U_0U_B^{-1}) = \begin{pmatrix}
        \mathbb{1}_d&0\\
        -B&\mathbb{1}_d
    \end{pmatrix} M_0 \begin{pmatrix}
        \mathbb{1}_d&0\\
        B&\mathbb{1}_d
    \end{pmatrix} ~,
\end{align}
where $U_0=U_B|_{B=0}$ and
\begin{align}\label{M0prod}
    M_0 = U_0 M_\complex U_0^{-1} = \begin{pmatrix}
        M_+&M_- G^{-1}\\
        GM_{-}\,\,&\,\,GM_{+}G^{-1}
    \end{pmatrix} ~, \qquad M_{\pm} \equiv \frac{M \pm \Bar{M}}{2} \,,
\end{align}
do not depend on the $B$-field. Moreover, because $B^T=-B$ the two lower-triangular matrices on the right-hand side of \eqref{MBproduct} are in $\text{O}(d,d;\reals)$. Hence, we are only left to prove that $M_0$ is an element of $\tO(d,d;\reals)$ by showing that $M_0^T J M_0 = J$. This can be easily verified by using the identities
\begin{equation}\label{usefulidentities}
    (M_-^TGM_+)^T=-M_-^TGM_+ \,, \qquad M_+^TGM_+ + M_-^T GM_- = G~,
\end{equation}
which, in turn, follow from \eqref{MGM}. Hence, the topological conditions \eqref{MGM} are equivalent to $M_B \in \text{O}(d,d;\reals)$. As a consequence, this implies that the Dirac quantization condition is preserved by the defect, namely $\mathbf{q}^T_L J \mathbf{q}_L = \mathbf{q}^T_R J \mathbf{q}_R \in 2\integers$.

The key observation is that, under certain rationality conditions on $G$, $B$, and $M_\complex$, we further have that $M_B\in\tO(d,d;\rationals)$. In particular,  $M_B$ admits a decomposition into a sequence of duality and gauging operations which leave $G$ and $B$ invariant. We can thus prove the existence of these topological defects in the quantum theory by combining these operations in half-space: the resulting theory has the same couplings $G$ and $B$ as the original one, and the corresponding $\tO(d)\times \tO(d)$ symmetry defect is topological.
We discuss how this is done in section \ref{section: topological defects from gauging and dualities}.

Before concluding this section, let us stress an important point. The correspondence between topological defects and their matrix action $M_B$ is not a bijection, as there are infinitely many distinct defects that correspond to the same $M_B$. However, defects that correspond to the same $M_B$ differ at most by fusing with condensation defects, i.e.~they differ by their kernel, while they act in the same way -- dictated by $M_B$ -- on the charges they do not annihilate. Indeed, condensation defects act as $M_B=\mathbb{1}_{2d}$ on the charges which are not in their kernel, namely on the charges which are not projected out by the (higher) gauging of the discrete subgroup of the zero-form symmetry.\footnote{In this sense, condensation defects -- which act as projectors -- can be intuitively thought of ``non-invertible identity operators."} Similarly, for a generic defect, the action on charges encoded in $M_B$ refers only to the charges which are not in the kernel (which is necessarily non-trivial if some of the transformed charges are not properly quantized). The non-invertibility of the defects reside precisely in having a non-trivial kernel, which is dictated by gauging and it is not visible at the level of the matrix $M_B$. However, once it is known how the defect is constructed in terms of gauging and dualities transformations, both the action on charges and its kernel are completely characterized. Notice that when $G$ and $B$ are such that $M_B \in \tO(d,d;\integers)$, we are actually describing self-dual points which enjoy ordinary zero-form symmetries which do not need any gauging. This generalizes the well-known fact that in the theory of a single compact scalar, T-duality at $R^2=1$ combines with reflection symmetry to produce an ordinary $\tO(1)\times\tO(1)\cong\integers_2 \times \integers_2$ zero-form symmetry.

\subsection{Topological defects as gauging and dualities}
\label{section: topological defects from gauging and dualities}
Our goal is to prove that the matrix $M_B$ implementing the action of topological defects on the charges as in \eqref{eq:action_on_Js} can be decomposed in a product of matrices corresponding to duality transformations and gauging of discrete subgroups, in the spirit of \cite{Choi:2021kmx,Choi:2022zal, Niro:2022ctq}, when certain rationality conditions are satisfied.
As we discussed in section \ref{section: topological defects from the stress tensor}, these operations act both on the couplings $G$ and $B$, as well as on the momentum and winding charges $\mathbf{q}$. The combination of such operations produces the transformation $M_B$, which acts on the charges but leaves the couplings invariant, as shown by \eqref{eq:action_on_Z}. In this section, we assume that $M_-$ defined in \eqref{M0prod} is an invertible matrix, and we comment on the case $\det M_- = 0$ in section \ref{detM-=0section}.\footnote{Note that when $d$ is odd either $M_+$ or $M_-$ have zero determinant, as it can be seen by \eqref{usefulidentities}, so that the discussion in this section applies only to the former case.}
We now prove the following claim.
\newline
\newline
{\it
Let $M_\pm = (M \pm \Bar{M})/2$, where $M$ and $\Bar{M}$ satisfy the topological condition \eqref{MGM}, and consider the action of the topological defect on charges, given by $M_B \in \tO(d,d;\reals)$ as in \eqref{MBproduct}. 
Assuming that $M_-$ is invertible, we have that $M_B \in \tO(d,d;\rationals)$ if and only if the following rationality conditions hold
\begin{align}\label{rationalityCond}
    M_+ + M_-G^{-1}B\,,\quad M_-G^{-1}\,,\quad GM_+G^{-1} - BM_-G^{-1} \quad \in \, \rationals^{d\times d}~.
\end{align}
Furthermore, when $M_B$ is rational, it can be decomposed as a finite sequence of gauging and $\tO(d,d;\integers)$ duality operations.
}
\newline
\newline
\indent To prove the first part, we explicitly compute $M_B$ from \eqref{MBproduct}. It is given by
\begin{align}\label{MBexplicit}
    M_B = \begin{pmatrix}
        M_+ + M_-G^{-1}B & M_- G^{-1}\\
        GM_{-} - BM_+ + (GM_{+}-BM_-)G^{-1}B&\quad (GM_{+}-B M_-)G^{-1}
    \end{pmatrix} \,.
\end{align}
Requiring that $M_B$ is rational is equivalent to requiring that its $(1,1)$, $(1,2)$, and $(2,2)$ blocks are rational, which in turns gives the conditions \eqref{rationalityCond}. Indeed, the $(2,1)$ block of $M_B$ is automatically rational because of the $\tO(d,d;\reals)$ condition and the invertibility of the $(1,2)$ block, which together imply
\begin{equation}\label{oddrelation}
 \begin{pmatrix}
        \alpha & \beta \\
        \gamma & \delta
    \end{pmatrix} \in \tO(d,d) \,, \quad
    \det\beta\neq 0 \quad    
    \quad \implies \quad \gamma = \beta^{-T}(\mathbb{1}_d - \delta^T\alpha)  \,.    
\end{equation}
Thus, under the assumption that $M_-$ is invertible, $M_B\in\tO(d,d;\rationals)$ if and only if the conditions \eqref{rationalityCond} are satisfied.

To prove the second part, it is convenient to define
\begin{equation}\label{LRKdef}
\begin{split}
L &= G M_+ M_-^{-1} - B \,, \\
R &= G M_-^{-1} M_+  + B \,, \\
K &= M_-^{-T}G \,.
\end{split}
\end{equation}
It is easy to show that $L$ and $R$ are antisymmetric, by using respectively the first of \eqref{usefulidentities} and its counterpart $(M_-G^{-1}M^T_+)^T=-M_-G^{-1}M^T_+$,
which follows from \eqref{MGM}. Notice that the rationality conditions \eqref{rationalityCond} can be equivalently reformulated as the condition that $L$, $R$, and $K$ are rational matrices. This will have a clear interpretation in the Lagrangian description of the topological defects, discussed in section \ref{defectLag}. Using the definitions \eqref{LRKdef}, we can rewrite $M_B$ as
\begin{equation}\label{MBExpansion}
M_B=\begin{pmatrix}
        K^{-T}R \quad& K^{-T} \\
        K+LK^{-T}R \quad & LK^{-T}
    \end{pmatrix} = J
    \begin{pmatrix}
        \mathbb{1}_d & L \\
        0 & \mathbb{1}_d
    \end{pmatrix} \begin{pmatrix}
        K & 0 \\
        0 & K^{-T}
    \end{pmatrix} \begin{pmatrix}
        \mathbb{1}_d & 0 \\
        R & \mathbb{1}_d
    \end{pmatrix}\,,
\end{equation}
where $K$ is rational and invertible, $L$ and $R$ are rational and antisymmetric. This implies that each factor in the decomposition above is an element of $\tO(d,d;\rationals)$. Next we prove that each factor in \eqref{MBExpansion} can be realized by duality and gauging operations.
\begin{itemize}
\item First, we consider
    \begin{align}\label{Qmatrix}
        \begin{pmatrix}
            \mathbb{1}_d&L\\
            0&\mathbb{1}_d
        \end{pmatrix}\in\tO(d,d;\rationals)~,
    \end{align}
where $L$ is antisymmetric and rational, namely
\begin{align}
    L_{ij} & = \frac{p_{ij}}{q_{ij}}~, \qquad \text{ with }\quad p_{ij}=-p_{ji} \in \integers \,, \quad q_{ij}=q_{ji} \in\integers_{\neq 0} \,.
\end{align}
Notice that $L$ can always be written as $\cG^{-1}\Tilde{L}\cG^{-1}$, where $\cG$ is of the form \eqref{gauging} and $\Tilde{L}$ is an antisymmetric integer matrix. Indeed, we can pick a diagonal invertible matrix $\cG$ with\footnote{If $\cG_{ii}<0$ for some $i=k$, the matrix $\cG$ is actually the product of a gauging operation with diagonal elements $|\cG_{ii}|$ and a duality element of the form \eqref{basischangematrix}, with $A$ diagonal such that $A_{ii}=-1$ for $i=k$ and $A_{ii}=+1$ for $i\neq k$. With a slight abuse of notation, we will refer to $\cG$ as a gauging operation even in this case.}
\begin{align}
    \mathcal G_{ii}  = \prod_{k \neq i} q_{ik}~, \qquad \text{ (no sum over }i) \,.
\end{align}
It is now straightforward to check that 
\begin{align}
    \Tilde{L}_{ab} \equiv (\cG L \cG)_{ab}  = p_{ab} \, q_{ab} \prod_{k\neq a,b}q_{ak}\,\prod_{\ell\neq a,b}q_{b\ell}\in\integers~,
\end{align}
so that $\Tilde{L}$ is integer and antisymmetric. All in all, we can write the matrix in \eqref{Qmatrix} as
\begin{align}
    \begin{pmatrix}
            \mathbb{1}_d&L\\
            0&\mathbb{1}_d
        \end{pmatrix} =
        \begin{pmatrix}
            \mathbb{1}_d\,&\,\mathcal G^{-1}\Tilde{L} \mathcal G^{-1}\\
            0&\mathbb{1}_d
        \end{pmatrix} =
        \begin{pmatrix}
        \mathcal G ^{-1}&0\\
        0&\mathcal G
    \end{pmatrix} \begin{pmatrix}
            \mathbb{1}_d&\Tilde{{L}}\\
            0&\mathbb{1}_d
        \end{pmatrix} \begin{pmatrix}
        \mathcal G&0\\
        0&\mathcal G^{-1}
    \end{pmatrix}~,
\end{align}
which is the combination of two gauging operations and an $\tO(d,d;\integers)$ duality element, as sought.

Given that $R$ is also an antisymmetric rational matrix, analogous arguments show that
\begin{align}
        \begin{pmatrix}
            \mathbb{1}_d&0\\
            R&\mathbb{1}_d
        \end{pmatrix}\,
\end{align}
can be written as a combination of dualities and gauging (this matrix can be put in the same upper-triangular form as in \eqref{Qmatrix} by conjugating it with $J$, which is itself a duality transformation).

\item Second, we consider
    \begin{align}\label{rationalM+M+-T}
        \begin{pmatrix}
            K&0\\
            0&K^{-T}
        \end{pmatrix}\in\tO(d,d;\rationals)~,
    \end{align}
where $K\in\text{GL}(d;\rationals)$.
This matrix can be expressed as a finite product of matrices of the form diag$(\cZ,\cZ^{-T})$, where $\cZ$ is one of the generating elements of $\text{GL}(d;\rationals)$.\footnote{This follows from the fact that any invertible matrix is the product of a finite number of matrices corresponding to elementary row operations, which are indeed the generating elements of $\text{GL}(d)$.} These are permutation matrices $\cP$ (i.e.~matrices that have exactly one entry of 1 in each row and each column, with all other entries being 0), elementary matrices $E_k(p/q)$ corresponding to multiplying a single $k^{\rm th}$ row of $\mathbb{1}_d$ by a rational number $p/q$,
\begin{align}
E_k(p/q) = \text{diag}\left(1,\dots,1,\frac{p}{q},1\dots,1\right)~,
\end{align}
and elementary matrices $T_{\ell,m}(p/q)$ corresponding to shifting the $\ell^{\rm th}$ row by a rational multiple $p/q$ of the $m^{\rm th}$ row,
\begin{align}
    T_{\ell,m}(p/q) & = \mathbb{1}_d + \frac{p}{q}e_{\ell,m}~, \qquad \text{ where } (e_{\ell,m})_{ij} = \delta_{i\ell} \delta _{jm} \text{ and } \ell\neq m\,.
\end{align}
Hence, it is sufficient to establish that each generating element $\cZ$ gives rise to a matrix diag$(\cZ,\cZ^{-T})$ which is a combination of dualities and gauging. This is obviously true for $\cP$ and $E_k$. The former correspond to duality elements, as $\text{diag}(\cP,\cP^{-T})$ is a generating element of $\tO(d,d;\integers)$. The latter correspond to gauging operations, as $\text{diag}(E_k, E_k^{-T})$ is of the form \eqref{gauging}. For $T_{\ell,m}(p/q)$, it is sufficient to establish that this property holds for $q=1$, as one can check that $T_{\ell,m}(p/q) =E_\ell(1/q) \, T_{\ell,m}(p) \, E_\ell(q)$. Since $T_{\ell,m}(p)\in\text{GL}(d;\integers)$, it follows that $\text{diag}(T_{\ell,m}(p),T_{\ell,m}(p)^{-T})$ is an $\tO(d,d;\integers)$ duality element.

Hence, we can obtain any matrix of the form \eqref{rationalM+M+-T}, with $K\in\text{GL}(d;\rationals)$, as a combination of gauging operations and duality elements, as sought.
    
\end{itemize}

\noindent This completes the proof that, when $M_B\in \tO(d,d;\rationals)$ or, equivalently, when the rationality conditions \eqref{rationalityCond} are satisfied, topological defects acting on the fields as the $\tO(d)\times \tO(d)$ elements in \eqref{eq:action_phi}, and on the charges as in \eqref{eq:action_on_charges}, can be obtained as a finite sequence of $\tO(d,d;\mathbbm Z)$ dualities and gaugings of non-anomalous discrete subgroups of the global symmetry \eqref{eq:symmetries}.
Individually, each duality or gauging operation acts on the couplings $G$ and $B$, but their combination leaves the couplings invariant (as it should be, since $M_B$ has this property).
By applying these operations in half-space, we have thus proven that such $\tO(d)\times \tO(d)$ topological defects exist at the quantum level and on a worldsheet $\Sigma$ of arbitrary topology.\footnote{It is useful to compare to the case of 4d Maxwell theory \cite{Niro:2022ctq}. There, the duality group is $\SL(2,\integers)$ and the non-invertible defects correspond to $U(1)$ elements acting on charges as $\SL(2,\rationals)$. The group $U(1)$ is the stabilizer of the $\tau=i$ element in $\SL(2,\reals)$. Here, the duality group is $\tO(d,d;\integers)$ and the non-invertible defects correspond to $\tO(d)\times\tO(d)$ elements acting on charges as $\tO(d,d;\rationals)$. The group $\tO(d)\times\tO(d)$ is the stabilizer of the $Z(G=\mathbb{1}_{d},B=0)=\mathbb{1}_{2d}$ element in $\tO(d,d;\reals)$.}

In section \ref{defectLag} we adopt a Lagrangian approach to describe the defects discussed in this section, and we show how the rationality conditions \eqref{rationalityCond} can be proven using an explicit Lagrangian realization of the defect action \eqref{eq:action_phi}.


\subsection{Comments on the case $\det M_- = 0$}
\label{detM-=0section}

In the previous section, we have proven that whenever $\det M_- \neq 0$, any matrix $M_B \in \tO(d,d;\rationals)$ can be decomposed as a sequence of duality and gauging operations. The assumption that $M_-$ is invertible simplifies both the rationality conditions (as it is enough that three out of the four blocks of $M_B$ are rational) and the task of providing an explicit decomposition of $M_B$ (as one block is invertible). The rationale behind this assumption can be understood from rewriting \eqref{eq:action_phi} in the basis of $d\phi$ and $i\star d\phi$: using the change of basis explained in footnote \ref{basischange}, it is easy to show that the topological defects act as
\begin{align}
    \begin{pmatrix}
        d{\phi}_L|_\gamma\\
        i\star d{\phi}_L|_\gamma
    \end{pmatrix} & = \begin{pmatrix}
        M_+&M_-\\
        M_- &M_+
    \end{pmatrix}\begin{pmatrix}
        d\phi_R|_\gamma\\
        i\star d{\phi}_R|_\gamma
    \end{pmatrix}~.
\label{actiononphistarphi}
\end{align}
Requiring that $\det M_- \neq 0$ is equivalent to requiring that one can express the transformation of $i\star d\phi_{L,R}$ in terms of $d\phi_{L,R}$.\footnote{This condition is analogous to the case $\sin\varphi \neq 0$ for the non-invertible defects of 4d Maxwell theory studied in \cite{Niro:2022ctq}.} This fact is crucial when giving a Lagrangian description of the defects in terms of a topological theory that can be written entirely in terms of $\phi_{L,R}$, which we propose in section \ref{defectLag}.

If $\det M_- \neq 0$ the rationality conditions are more involved, because in order to have $M_B \in \tO(d,d;\rationals)$ we need to require that all of its four blocks, independently, are rational matrices. Moreover, the explicit decomposition given in the previous section does not apply. However, we believe that even in this more general case any $M_B \in \tO(d,d;\rationals)$ can be decomposed as a (finite) sequence of dualities and gaugings. Notice that this is straightforward to prove if any matrix of $\tO(d,d;\rationals)$ can be written as a product of a finite number of elements of the form \eqref{Bshiftmatrix}, \eqref{basischangematrix}, and $\eqref{generalizedTmatrix}$, with $\Theta$ being an antisymmetric rational matrix and $A \in \text{GL}(d,\rationals)$.\footnote{Notice that the decomposition \eqref{MBExpansion}, valid in the case $\det M_-\neq 0$, features only elements of the form \eqref{Bshiftmatrix} and \eqref{basischangematrix}, up to conjugation with $J$.}

In the rest of this section, we analyze in the detail the case $M_-=0$, where the topological conditions imply that $M_+$ is invertible, as it satisfies
\be
M = \Bar{M} = M_+ \,, \qquad M_+^T G M_+ = G \,.
\label{M+forM-=0}
\ee
Notice that the scalars $\phi^i$ are not mixed with their duals $\Tilde{\phi}^i$ in this case.
The action on charges $M_B$ can be decomposed as
\begin{align}
    M_B = \begin{pmatrix}
        M_+ & 0\\
        - BM_+ + GM_{+}G^{-1}B&\quad GM_{+}G^{-1}
    \end{pmatrix} =
    \begin{pmatrix}
        M_+ & 0\\
        0 & M_+^{-T}
    \end{pmatrix}
    \begin{pmatrix}
        \mathbb{1}_d & 0\\
        -\Delta &\quad \mathbb{1}_d
    \end{pmatrix} \,,
\label{MBforM-0}
\end{align}
where $\Delta$ is the antisymmetric matrix
\be
\Delta = M_+^T B M_+ - B = - \Delta^T \,.
\label{deltadefinition}
\ee
The rationality conditions simply read
\be
M_+ \,, \quad \Delta \quad \in \rationals^{d\times d} \,,
\ee
so that they do not involve $G$. We are thus describing a family of symmetries that exist for any (real) metric $G$. The most familiar case is $M_+=-\mathbb{1}_d$ (so that $\Delta=0$), which corresponds to an invertible reflection zero-form symmetry acting as $\phi^i \rightarrow -\phi^i$, existing for any $G$ and $B$. A generic $M_+$ is a $G$-orthogonal rotation (due to \eqref{M+forM-=0}), acting as $\phi^i \rightarrow (M_+)^i_j \phi^j$ -- notice that for $G=\mathbb{1}_d$ we have $M_+ \in \tO(d;\rationals)$. If we were to consider just how such rotation transforms the action \eqref{eq:compact_action}, we would get
\begin{align}
    S[\phi] \rightarrow S'[\phi]=\frac{1}{4\pi} \int_\Sigma \left( G_{ij}\, d\phi^i\wedge\star d\phi^j + i (B_{ij}+\Delta_{ij})\, d\phi^i\wedge d\phi^j \right) \,,
\label{actiontransformM-=0}
\end{align}
where we have used \eqref{M+forM-=0} and the definition \eqref{deltadefinition}. This is not an ordinary symmetry because it acts by shifting the $B$-field by $\Delta$, so it must include a transformation of the type $\eqref{Bshiftmatrix}$ with $\Theta=\Delta$, which is precisely the right matrix in the decomposition \eqref{MBforM-0}.\footnote{Recall that $\cM_\Theta$ in \eqref{Bshiftmatrix} acts on charges as $\mathbf{q} \rightarrow \cM_\Theta^{-T} \mathbf{q}$.} This is combined with the left matrix in the decomposition \eqref{MBforM-0}, which is a transformation of the type $\eqref{basischangematrix}$ with $A=M_+^{-T}$, such that their combination leaves both $G$ and $B$ invariant.

All in all, we have found that in the theory of $d$ compact scalars, a $G$-orthogonal rotation is (generically) a non-invertible symmetry, present for rational values of $\Delta$ and of the rotation matrix $M_+$, but for any metric $G$.
In section \ref{lagrangianfusion} we give an explicit Lagrangian description of a defect which corresponds to $M_-=0$, by considering the fusion of two identical defects.

We do not attempt here to provide an explicit decomposition of $M_B$ in the generic case where $M_-$ has rank $0<r<d$. However, it is conceivable that such a case corresponds, in a suitable basis which we can reach by using matrices of the form \eqref{generalizedTmatrix}, to transformations that do not mix $d-r$ scalars with their duals -- hence, they are described by rotations in a $(d-r)\times(d-r)$ version of what happens in the $M_-=0$ case -- and act on the other $r$ scalars and their duals as in the $\det M_- \neq 0$ case -- hence, they are described by an $r \times r$ version of the action on charges decomposed as in \eqref{MBExpansion}.


\subsection{Fusion rules}
\label{fusionrulessection}

In the previous sections, we proved the existence of topological defects on the worldsheet at the quantum level, when the metric $G$ and the $B$-field satisfy the rationality conditions \eqref{rationalityCond}. We achieved this by showing that any such defect acts on charges as a rational matrix $M_B$, which can be realized by a sequence of discrete gaugings and duality transformations. In this section, we discuss the fusion rules of these defects.

A generic defect in \eqref{MBExpansion} can be realized by a finite series of transformations
\begin{equation}
    \cM_1\mathfrak{g}_1 \cM_2\mathfrak{g}_2  \dots \cM_n\mathfrak{g}_n \,,
    \label{seriestransformations}
\end{equation}
where $\cM_i \in \tO(d,d;\mathbbm{Z})$ are duality transformations and $\mathfrak{g}_i=\text{diag}(\mathcal{G}_i^{-1},\mathcal{G}_i)$ correspond to gaugings of discrete subgroups of the zero-form symmetry \eqref{eq:symmetries}. Each component of the form $\cM\mathfrak{g}$ is an interface between two different theories, as it maps the theory to another one with different couplings, but the whole sequence in \eqref{seriestransformations} is a topological defect which leaves the couplings invariant. 
In the following, we discuss the fusion of one of such interfaces. Moreover, these fusion rules can be used to determine the fusion of the topological defects of the previous sections.

\begin{figure}
\centering
    \begin{tikzpicture} 
\draw[] (2,0) -- (2,3);
\draw[] (4,0) -- (4,3);
\filldraw[color=white, opacity = 0.4, left color =gray!70, right color =white] (4,0) rectangle (7,3);
\node[anchor = north] at (2,0) {$\cM$};
\node[anchor = north] at (4,0) {$\mathfrak{g}$};
\end{tikzpicture}
    \caption{On the right space (gray area) we have gauged a discrete subgroup of the zero-form symmetry with Dirichet boundary conditions, leading to the $\mathfrak{g}$ transformation. The interface on the left implements a duality transformation $\cM$.}
    \label{fig:slab}
\end{figure}
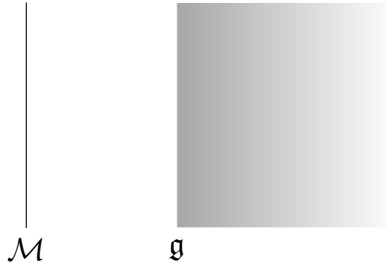

Let us consider the action of an interface implementing the transformation $\cM \frak{g}$ on a vertex operator labeled by the charge vector $\mathbf{q}$ taking values in a charge lattice $\Lambda$.
This action can be determined by applying the action of $\cM$ and $\frak{g}$ in steps.
We can view the interface  corresponding to $\cM\mathfrak{g}$ as a combination of two interfaces \cite{Kaidi:2021xfk, Choi:2021kmx} as in figure \ref{fig:slab}. The first interface implements a duality transformation mapping the theory to a dual one, while on the right-hand side of the second interface we have gauged a discrete subgroup of the zero-from symmetry.  We can obtain the interface $\cD_{\cM\mathfrak{g}}$ by collapsing the slab between these two interfaces.

Consider now dragging a vertex operator from the right to the left through the two interfaces in figure \ref{fig:slab}. First, the interface implementing $\frak{g}$ acts as a projector onto invariant states under the discrete group that is being gauged. Hence, if the resulting vertex operator is gauge-invariant it will remain a local operator, while if the resulting operator is not gauge-invariant it will be attached to an invertible line. Then, the duality interface acts as a duality transformation

\begin{equation}
  V_{\mathbf{q}} \rightarrow   e^{i\theta_\cM(q)}V_{\mathbf{\cM q}} \,,
\end{equation}
where $\theta_\cM(\mathbf{q})$ is a phase which we determine in appendix \ref{app:phase}. Thus, we conclude that an interface $\cD_{\cM\mathfrak{g}}$ realizing the transformation $\cM\mathfrak{g}$ acts on a vertex operator $V_\mathbf{q}$ as

\begin{equation}\label{eq:actionV}
  V_{\mathbf{q}} \rightarrow
 \begin{cases}
 e^{i\theta_\cM(\mathbf{q})} V_{\cM\mathfrak{g} \mathbf{q}} & \text{ if } \frak{g} \, \mathbf{q} \in \Lambda \,,
\\
\text{non-local primary} & \text{ otherwise } .
\end{cases} 
\end{equation}
When acting on a state, we additionally need to multiply by the quantum dimension $\dim \cD_{\cM \frak{g}}$ of the interface. Thus, the interface $\cD_{\cM\mathfrak{g}}$ realizing the transformation $\cM\mathfrak{g}$ acts on the state $\ket{\mathbf{q}}$ as

\begin{equation}\label{eq:actionStates}
 \cD_{\cM\mathfrak{g}} \ket{\mathbf{q}} = \begin{cases}
 \dim \cD_{\cM \frak{g}} \,\, e^{i\theta_\cM(\mathbf{q})}\ket{ \cM \frak{g}\; \mathbf{q}} & \text{ if } \frak{g} \, \mathbf{q} \in \Lambda\,,
\\
0 & \text{ otherwise } \,.
\end{cases} 
\end{equation}

The orientation-reversal of the interface, denoted as $\Bar\cD_{\cM \frak{g}}$, generates the transformation $\mathfrak{g}^{-1} \cM^{-1}$ on the charges. The matrix $\mathfrak{g}$ is obtained by gauging the discrete subgroup $\widehat{G}=\prod_i \mathbbm{Z}_{p_i}$, which is a direct product of Abelian groups.
We thus see that the fusion rule of $\cD_{\cM \frak{g}}$ with its orientation-reversal takes the form 
\begin{equation}
\label{eq:fusion}
\bar\cD_{\cM \frak{g}} \otimes \cD_{\cM \frak{g}} = \prod_i\sum\limits_{k_i=0}^{p_i-1} \exp\left(\frac{2\pi i}{p_i} k_i \oint  v_q\right) ~,    
\end{equation}
where $v_q$ is defined in \eqref{def_vq}.
The operator on the right-hand side is a projector onto $\widehat{G}$-invariant states.
Notice that the above fusion rule is non-Abelian: if we consider the product $\cD_{\cM \frak{g}} \otimes \bar \cD_{\cM \frak{g}}$ the condition on the right-hand side of \eqref{eq:actionStates} is different. From \eqref{eq:fusion} we see that the quantum dimension of $\cD_{\cM \frak{g}}$ is
\begin{equation}
    \dim \cD_{\cM \frak{g}} = \prod_i \sqrt{p_i}~.
\end{equation}
Here we only determined the fusion of $\cD_{\cM \frak{g}}$ with its orientation-reversal, but for any two different elements one can use \eqref{eq:actionStates} to determine their fusion. Similarly, once the decomposition \eqref{seriestransformations} is determined for two generic topological defects, one can determine their fusion using the method discussed in this section.

\section{Lagrangian description of the topological defects}
\label{section: Lagrangian description of the topological defects}
In this section, we give an explicit Lagrangian description of the topological defects. This allows to prove the rationality conditions \eqref{rationalityCond} and to compute the fusion rules.

\subsection{Defect Lagrangian and rationality conditions}
\label{defectLag}
As we discussed in section \ref{section: topological defects from the stress tensor}, we consider a defect that splits the worldsheet $\Sigma$ as $\Sigma=\Sigma_L \cup \Sigma_R$. We can write the total action in terms of the left and the right fields $\phi_{L,R}$, as the sum of the bulk contributions $S[\phi_L]$ and $S[\phi_R]$ as in \eqref{eq:compact_action} (with the same couplings $G$ and $B$) and a topological defect term localized on $\gamma$,
\begin{multline}
\label{Lagrangian action}
S = \int_{\Sigma_L}  \left( \frac{G_{ij}}{4\pi} d\phi_L^i \wedge \star d\phi_L^j + \frac{iB_{ij}}{4\pi} d\phi_L^i \wedge d\phi_L^j \right) + \int_{\Sigma_R} \left( \phi_L \leftrightarrow \phi_R \right) \\
+ \int_{\gamma}  \left( \frac{iL_{ij}}{4\pi}\phi_L^i \wedge d\phi_L^j +  \frac{iR_{ij}}{4\pi}\phi_R^i \wedge d\phi_R^j +  \frac{iK_{ij}}{2\pi} \phi_L^i \wedge d\phi_R^j \right) \,,
\end{multline}
where the matrices $L$ and $R$ are antisymmetric. The defect couplings $L$, $R$, and $K$ are \textit{a priori} independent from the quantities defined in \eqref{LRKdef}, but we show below that they are identified after imposing the equations of motion.

Note that the defect Lagrangian does not depend on the worldsheet metric, and hence this term does not contribute to the stress tensor. In principle, the compactness of the fields $\phi_{L,R}$ requires the matrices $L$, $R$, and $K$ to have integer entries. However, by coupling the above Lagrangian to compact scalar fields localized on the defect $\gamma$, it is possible to effectively make those couplings rational. Naively, this can be seen by integrating out the defect fields. More precisely, one should consider the equations of motion arising from the full action (including the defect fields) and isolate the equations for the fields $\phi_{L,R}$, as explained for the $d=1$ case in \cite{Niro:2022ctq}. This produces the same result as starting from the ``effective action" \eqref{Lagrangian action} with rational defect couplings $L$, $R$, and $K$.\footnote{This is analogous to the topological Lagrangian that describes the fractional quantum Hall effect: the Chern-Simons level is effectively rational, and it characterizes the fractional response of the system arising from a well-defined and gauge-invariant action which includes the coupling to an auxiliary dynamical field.}

Imposing the variational principle, we get both the usual bulk equations of motion $d\star d \phi^i_L = d\star d \phi^i_R = 0$ and the defect conditions relating left and right fields on $\gamma$ as
\begin{equation}
\begin{split}
\left(-iG_{ij} \star d\phi^j_L + (B_{ij}+L_{ij})d\phi_L^j + K_{ij}d\phi^j_R\right)\big|_{\gamma} &= 0 \,,\\
\left(-iG_{ij} \star d\phi^j_R + (B_{ij}-R_{ij})d\phi_R^j + K_{ji}d\phi^j_L\right)\big|_{\gamma} &= 0 \,.
\end{split}
\end{equation}
These relations encode the action of the defect (which we take at $x=0$ for simplicity) on the fields. In complex coordinates and using matrix notation they read
\begin{equation}
\begin{split}\label{defectBC}
(-G+B+L)\partial \phi_L + (-G-B-L)\bar\partial \phi_L &= -K \partial \phi_R + K \bar\partial \phi_R \,, \\
(-G+B-R)\partial \phi_R + (-G-B+R)\bar\partial \phi_R &= -K^T \partial \phi_L + K^T \bar\partial \phi_L \,,
\end{split}
\end{equation}
where it is understood that these equations hold along $\gamma$.

As established in (\ref{MGM}), topological defects do not mix holomorphic with antiholomorphic components and they relate left and right fields as the $\tO(d;\reals)\times\tO(d;\reals)$ transformation
\begin{equation}\label{defectaction}
\partial \phi_L = M \partial \phi_R \,,  \quad \bar\partial \phi_L = \bar{M} \bar\partial \phi_R \,,
\end{equation}
where $M$ and $\bar M$ are independent real matrices satisfying the topological condition $M^T G M = \Bar{M}^T G \Bar{M}=G$. Plugging \eqref{defectaction} in \eqref{defectBC}, we get the four matrix equations which relate the defect action $M$ and $\Bar{M}$ to the defect Lagrangian data $L$, $R$, and $K$ for a theory with fixed coupling $G$ and $B$. They are
\begin{equation}
\begin{split}\label{set2}
2G &=K (M^{-1} -\bar{M}^{-1}) \,,\\
2G &=K^T (M-\bar{M}) \,,\\
2(L+B) &=-K (M^{-1}+\bar{M}^{-1}) \,,\\
2(R-B) &=+K^T (M+\bar{M})\,. \\
\end{split}    
\end{equation}
Recalling the definition of $M_\pm = (M \pm \Bar{M})/2$, it follows from the second equation that $\det K, \det M_{-} \neq 0$. Notice that the condition of $M_-$ being invertible (which we assumed in section \ref{section: topological defects from gauging and dualities}) arises naturally in the Lagrangian approach.\footnote{Indeed, to satisfy the variational principle, the boundary variation of the action coming from the bulk Lagrangian needs to be canceled by the one coming from the defect Lagrangian. As the former depends on the Hodge star of the fields, whereas the latter does not, this cancellation can be achieved only if one can express $i\star d \phi_{L,R}$ in terms of $d\phi_{L,R}$. It is easy to show, by using \eqref{actiononphistarphi}, that this is possible if and only if $M_-$ is invertible.} Hence, the second equation is solved by
\begin{equation}
\label{M-solution}
M_- = K^{-T} G \,.    
\end{equation}
Using the topological condition for $M$ and $\Bar{M}$, it follows that this also solves the first equation. The third and the fourth equations are respectively solved by
\begin{equation}
\label{M+solution}
M_+= G^{-1}(L+B)K^{-T}G \,, \qquad M_+ = K^{-T}(R-B) \,,    
\end{equation}
which imply that the defect couplings must satisfy the constraint
\begin{equation}
\label{LRKconstraint}
G^{-1}(L+B)K^{-T}G = K^{-T}(R-B) \,,    
\end{equation}
which allows to express $L$ (or $R$) in terms of the other matrices. Moreover, by combining the equations of motions we also get the constraint
\begin{equation}
\label{secondconstraint}
G K^{-T} G + (B+L) K^{-T} (B-R) = K \,.    
\end{equation}
Equations \eqref{M-solution}-\eqref{secondconstraint} are equivalent to the equations of motion supplemented with the topological conditions, and they exhaust all the independent relations that follow from the Lagrangian.

To summarize, given the Lagrangian description in \eqref{Lagrangian action} in terms of the bulk couplings $G$ and $B$, and of the rational defect couplings $L$, $R$, and $K$ -- satisfying $L^T=-L$, $R^T=-R$, $\det K \neq 0$, and the constraints \eqref{LRKconstraint} and \eqref{secondconstraint} -- the defect acts on the fields as in \eqref{defectaction} with $M_-$ and $M_+$ given by \eqref{M-solution} and \eqref{M+solution}, respectively, such that $M$ and $\Bar{M}$ satisfy the topological conditions \eqref{MGM}.

We can easily invert \eqref{M-solution} and \eqref{M+solution} to express $K$, $L$, and $R$ in terms of $G$, $B$, and $M_\pm$. The result is precisely \eqref{LRKdef}.
Moreover, one can verify that the relation \eqref{secondconstraint} is equivalent to imposing that the $(2,1)$ block of $M_B$ as in \eqref{MBExpansion}, namely $K+LK^{-T}R$, is given by the correponding combination in \eqref{MBexplicit}. This proves that the defect action on charges constructed with the Lagrangian matches the one we constructed in section \ref{section: Topological defects on the worldsheet}, and it is an element of $\tO(d,d;\reals)$.
Moreover, notice that the rationality of the defect couplings $K$, $L$, and $R$ is equivalent to the rationality conditions \eqref{rationalityCond}. This implies that in fact $M_B \in \tO(d,d;\rationals)$ within the Lagrangian approach.

\subsection{Independent data of the defects}

Let us now count the number of independent (rational) data of the defects, for a fixed choice of the bulk couplings $G$ and $B$, which consist of $d^2$ independent parameters. These can be counted with the Lagrangian approach, as the number of independent components of $L$, $R$, and $K$, subject to the constraints \eqref{LRKconstraint} and \eqref{secondconstraint}. Equivalently, in the approach of section \ref{section: Topological defects on the worldsheet}, this is the number of independent components of $M_+$ and $M_-$ subject to the topological condition \eqref{MGM}, or also the number of independent components of $M_B$ which are not fixed by the bulk couplings.

Let us start with the Lagrangian approach. The constraint \eqref{LRKconstraint} allows us to express $R$ in terms of $L$ and $K$. Then, the constraint in 
\eqref{secondconstraint} leads to
\begin{equation}
G = (B+L)G^{-1}(B+L) + K G^{-1} K^T \,.    
\end{equation}
These are $\frac{d(d+1)}{2}$ independent constraints, as $G$ is symmetric. The number of independent (rational) parameters of the defects is then
\begin{equation}
\left( d^2 + \frac{d(d-1)}{2} \right) - \frac{d(d+1)}{2} = d(d-1) \,.    
\end{equation}

The same result can be obtained with the approach of section \ref{section: Topological defects on the worldsheet}. The number of parameters in $M_+$ and $M_-$ is $2d^2$, but they need to satisfy the two symmetric topological conditions \eqref{MGM}, which reduce the number of independent (rational) parameters of the defects to
\begin{equation}
2d^2 - 2\left(\frac{d(d+1)}{2}\right)  = d(d-1) \,,    
\end{equation}
consistently with the previous counting. Clearly, this is the same as the number of independent components of the two $G$-orthogonal matrices $M$ and $\Bar{M}$, which is precisely the dimension of the group $\tO(d)\times\tO(d)$. Alternatively, the number of independent components of the matrix $M_B \in \tO(d,d)$ specifying the action of the defect on charges is $d(2d-1)$. However, $d^2$ of them are given in terms of the bulk data $G$ and $B$, so we are again left with $d(d-1)$.

In the simple case $d=1$ (where $G=\cR^2$ and $B=0$),\footnote{To avoid confusion between the matrix of right couplings and the radius of the compact scalar in the $d=1$ case, we denote the latter -- only in this subsection -- by $\mathcal{R}$.} this gives zero independent parameters. This is expected, as $L=R=0$, so that $K \in \rationals$ is the only (rational) data of the defect. However, while \eqref{LRKconstraint} is trivially satisfied, the constraint \eqref{secondconstraint} imposes $K=\pm \cR^2$, and no free parameter is left. This implies that these defects can be constructed via duality and gauging only for $\cR^2\in\rationals$. Moreover, the fact that there are no free parameters means that there is only (up to fusing with condensation defects) a \text{finite} number of defects, and in particular only two defects exist: the $\cT$-defect and its reflection-conjugate, which correspond to $K=\pm \cR^2$. 
This is consistent with the Lagrangian analysis of the $d=1$ case discussed in \cite{Niro:2022ctq}.
The same result can be obtained by noticing that the solutions \eqref{M-solution} and \eqref{M+solution} in $d=1$ require $M_-=\cR^2/K$ and $M_+=0$. As $K=\pm \cR^2$, no free parameter is left and only the two defects with $M_+=0$ and $M_-=\pm 1$ exist.
Notice that in the Lagrangian approach we are working in the case where $M_-$ is invertible, so the counting of the defects does not include the trivial and the reflection defects, corresponding to $M_+=\pm 1$ and $M_-=0$. They generate an ordinary $\integers_2$ symmetry and they exist for any value of $\cR^2$. These defects are correctly found if we use instead those results of section 2 which hold regardless of the assumption that $M_-$ is invertible. Indeed, the topological conditions for $M_+$ and $M_-$ for $d=1$ have four solutions, corresponding to the four elements $(M_+,M_-)=(\pm 1, \pm 1)$ of $\tO(1)\times\tO(1)\cong\integers_2\times\integers_2$, and $\cR^2 \in \rationals$ follows from requiring that the rationality conditions hold.

For $d>1$ we have instead a non-zero number of independent defect parameters, which implies that for fixed bulk couplings $G$ and $B$, there is an infinite number of $\tO(d)\times\tO(d)$ defects, corresponding to arbitrary choices of the $d(d-1)$ rational parameters. Ultimately, the peculiarity of the $d=1$ case is related to the fact that $\tO(d)\times\tO(d)$ is a finite (Abelian) group if and only if $d=1$.

\subsection{Fusion rules from the Lagrangian}
\label{lagrangianfusion}
Let us now see how the fusion rules of two defects can be determined with the Lagrangian approach, focusing on the fusion of a defect with itself (but the method is general and can be applied for any two defects with a Lagrangian description).

In the $d=1$ case, it is well-known that the $\cT$-defect at $\cR^2=N$ squares to the sum of the $N$ generators of the $\integers_N$ subgroup of the momentum symmetry. For generic $d$, we consider fusing two identical defects $\cD[\gamma]$ and $\cD[\gamma']$ described by the Lagrangian \eqref{Lagrangian action}, by taking the limit where $\gamma'\rightarrow\gamma$ so that the intermediate region between them shrinks to zero. In this way, the field $\phi_I$ in the intermediate region actually plays the role of a defect field. The action describing the fusion is given by the sum of the two left and right bulk terms as in \eqref{Lagrangian action}, and the defect contribution
\begin{multline}
S_{\cD^2} = \int_{\gamma}  \bigg( \frac{iL_{ij}}{4\pi}\phi_L^i \wedge d\phi_L^j +  \frac{iR_{ij}}{4\pi}\phi_R^i \wedge d\phi_R^j \\ +  \frac{i(L_{ij}+R_{ij})}{4\pi} \phi_I^i \wedge d\phi_I^j +  \frac{i}{2\pi} (K_{ij} \phi_L^i - K_{ji} \phi_R^i ) \wedge d\phi_I^j  \bigg) \,.
\end{multline}
For simplicity, let us focus on the case where $L=R=0$. The defect boundary conditions and the equation of motion for $\phi_I$ read
\bea
\left(+G_{ij} \star d\phi_L^j + i B_{ij} d\phi_L^j + i K_{ij} d\phi_I^j\right)\big|_{\gamma} &= 0 \,, \\
\left(-G_{ij} \star d\phi_R^j - i B_{ij} d\phi_R^j - i K_{ji} d\phi_I^j\right)\big|_{\gamma} &= 0 \,, \\
\left(K_{ij} d\phi_L^i - K_{ji} d\phi_R^i\right)\big|_{\gamma} &= 0 \,,
\label{boundaryconditionfusion}
\eea
which imply that $\cD^2$ acts on the charges as the $\tO(d,d;\rationals)$ transformation
\be
\mathbf{q}_L = M_B \, \mathbf{q}_R \,, \qquad M_B = \begin{pmatrix}
K^{-T} K & 0 \\ 0 & K K^{-T}    
\end{pmatrix} \,.
\ee
In the language of section \ref{detM-=0section}, this transformation corresponds to the case $M_-=0$, $M_+=K^{-T} K$, and $\Delta=0$.
Indeed, as $\det M_- =0$, the corresponding defect is not described by a Lagrangian of the form \eqref{defectaction}, but it requires a defect field $\phi_I$ which acts as a generalized Lagrange multiplier. The defect specified by \eqref{boundaryconditionfusion} is, for a generic choice of $K$, non-invertible.

A simple example is where $K$ is diagonal with elements $K_{ii}=p_i/q_i$, with $p_i$ and $q_i$ positive co-prime integers. We have that $M_B=\mathbb{1}_{2d}$, so that the defect acts trivially on all charges that are not in its kernel. Indeed, the defect Lagrangian simply reads
\begin{equation}
S_{\cD^2} = \sum_{i=1}^d \int_{\gamma}  \frac{i p_i}{2\pi q_i} (\phi_L^i - \phi_R^i ) \wedge d\phi_I^i \,,    
\end{equation}
and using \eqref{boundaryconditionfusion} we have that on the defect $\gamma$ left and right currents are related as
\be
(\star j^L_w)^i = (\star j^R_w)^i \,, \qquad 
(\star j^L_m)_i = (\star j^R_m)_i = \frac{p_i}{q_i} \frac{d\phi^i_I}{2\pi}\,.
\ee
The second equation implies that only momentum charges of the form $m_i=p_i/q_i n$, with $n\in \integers$, are acted on trivially by the defect, whereas the others are in its kernel. This corresponds to gauging, for each $i$, a $\integers_{p_i} \times \integers_{q_i}$ subgroup of the $\U(1)^{(i)}_m \times \U(1)^{(i)}_w$ symmetry along the defect, which thus acts as a projector onto $\Hat{G}$-invariant states, where $\Hat{G}=\prod_i \integers_{p_i} \times \integers_{q_i}$. This projector leads to the constraints we found before in \eqref{eq:actionV} by analyzing the fusion of defects on general grounds. Finally, notice that if $d=1$ the discussion of this section implies that the $\cT$-defect at $R^2=p/q \in \rationals$ squares to a condensation defect for the $\integers^{(m)}_{p} \times \integers^{(w)}_{q}$ non-anomalous subgroup of the zero-form symmetry. As this is not the identity operator (unless $p=q=1$, where one has an invertible $\integers_2$ $\cT$ symmetry at $R^2=1$), the $\cT$ symmetry is non-invertible.

\section{Applications to bosonic string theory}
\label{section: defects at d=1}

In the previous sections, we have shown that the worldsheet theory admits topological defects for special values of $G$ and $B$, namely when rationality conditions are satisfied. These defects are associated to symmetries that act on charges with a rational transformation $M_B$ and, as such, they are generically non-invertible, except for the self-dual points where $M_B \in \tO(d,d;\integers)$. It is natural to wonder about the fate of these defects in the $D$-dimensional target spacetime \cite{Bachas:2012bj}.

In this section, we focus for simplicity on a single compact direction of the target spacetime \cite{Fuchs:2007tx, Thorngren:2021yso, Roumpedakis:2022aik}. In this case $d=1$ (so that $B=0$), and there is a single non-invertible defect (up to a $\mathbbm{Z}_2$ reflection) for rational values of $G=R^2$, obtained by combining standard T-duality and gauging of a non-anomalous subgroup of the $\U(1)_m\times \U(1)_w$ global symmetry.

Indeed, consider a free compact scalar $\phi$ at radius $R^2 = p/q$ with $p$ and $q$ positive integers such that $\gcd(p,q)=1$. After T-duality, we can map the theory to the one with $\Tilde{R}^2=1/R^2=q/p$. Gauging the non-anomalous $\mathbbm{Z}_q \times\mathbbm{Z}_p$ subgroup of $\U(1)_m\times \U(1)_w$ acts on the radius as $\Tilde{R}\rightarrow (p/q)\Tilde{R}$. Therefore, the combination of the two operations leaves the radius invariant and, as such, it defines a topological defect. This is non-invertible away from the self-dual point $R=1$, and we will refer to it (and to the corresponding symmetry) as the $\cT$-defect.
 
To further simplify the discussion, we focus on the case where $R^2 = N$ is a positive integer. First, we analyze how the $\cT$ symmetry acts on the vertex operators corresponding to states in the target-space effective theory. This determines the action on the fields that appear in the $D$-dimensional spacetime effective action and we show that it leads to a non-invertible symmetry.
Then, we analyze the selection rules on worldsheets of different topology, and we derive them on the sphere and on the torus. Focusing on a particular one-point function, we verify the selection rules by a direct calculation.

\subsection{String theory on $\cM_{D-1}\times S^1$}

Consider the critical bosonic string (i.e.~$D=26$) compactified on $S^1$. We will use the index notation $I = (\mu,25)$, where $\mu,\nu=0, 1, \dots, 24$ label the non-compact directions, while the 25$^{\rm th}$ direction is compact. The worldsheet fields are denoted by $X^I = (X^\mu, \phi)$, with $\phi\sim\phi+2\pi$. Following the machinery of section \ref{section: Topological defects on the worldsheet}, we can define the $\cT$-defect at $R^2=N$ which acts on $\phi$ and on the charges $(w,m)$ as the one-dimensional version of equations \eqref{eq:actiononv_C} and \eqref{eq:action_on_charges}, namely
\begin{align}\label{1dtr}
    \begin{pmatrix}
        \partial\phi\\
        \Bar{\partial}\phi
    \end{pmatrix}
    \rightarrow \begin{pmatrix}
        1&0\\
        0&-1
    \end{pmatrix}\begin{pmatrix}
        \partial\phi\\
        \Bar{\partial}\phi
    \end{pmatrix}\,, \qquad
        \begin{pmatrix}
        w\\
        m
    \end{pmatrix}
    \rightarrow \begin{pmatrix}
        0&1/N\\
        N&0
    \end{pmatrix}\begin{pmatrix}
        w\\
        m
    \end{pmatrix}\,.
\end{align}
For the purposes of the following discussion, we restrict our attention to local operators that transform into other local operators under the action of the $\cT$ symmetry, and not to local operators which are mapped into non-local ones. The latter case will be addressed in due course. 

A $\cT$-defect defined on a closed curve $\gamma$ acts on local operators defined at a point $z\in\text{Int }\gamma$ by shrinking the contour $\gamma$ down to the point $z$. This process is topologically equivalent to dragging the defect across the insertion, which transforms the local operator into another local operator (by assumption), and then we are left with the transformed local operator at $z$ which does not link anymore with the defect. The latter can be thus smoothly shrunk to a point, producing the identity operator up to a factor of its quantum dimension $\sqrt{N}$ \cite{Thorngren:2019iar}. Using the state-operator map, the fields $(\partial\phi,\Bar{\partial}\phi)$ can be represented as states in the Hilbert space in the sense of radial quantization. Hence, the above discussion combined with (\ref{1dtr}) allows us to deduce the action of the $\cT$-defect on states as
 \begin{align}\label{TV}
    \cT \ket{\partial \phi}  = \sqrt{N}\ket{\partial \phi}\,, \qquad
    \cT \ket{\bar\partial \phi}  = - \sqrt{N}\ket{\bar\partial \phi} \,.
 \end{align}
Whenever the action of the $\cT$-defect on a local operator $\cO(z, \Bar{z})$ leads to a non-genuine operator, i.e.~to an operator that needs to be attached to a line to be well-defined, the corresponding state $\ket{\cO}$ is mapped by the $\cT$-defect to zero in the Hilbert space. The non-trivial kernel of $\cT$ characterizes its non-invertible nature.

Finally, we recall the fusion of the $\cT$-defect with itself for $R^2=N$, given by \cite{Fuchs:2007tx, Thorngren:2021yso}
\begin{align}\label{fusionruleR2=N}
    \cT(\gamma)\otimes\cT(\gamma) = \bigoplus_{u=0}^{N-1}\eta^u(\gamma) ~,
\end{align}
where $\gamma$ is a closed curve, and 
\begin{equation}\label{etadefinition}
    \eta(\gamma)  = \exp\left(\frac{2\pi i}{N}\int_\gamma \frac{d\Tilde\phi}{2\pi}\right) = \exp\left(-\int_\gamma\star d\phi\right)~
\end{equation}
is the defect that generates the $\integers_{N}$ subgroup of the $\U(1)_m$ symmetry, and $\eta^N=1$. We have used that the dual scalar $\Tilde{\phi}$ is defined as in \eqref{eq:momentumJ},
\begin{align}
d\Tilde{\phi} = i R^2 \star d\phi \,.
\end{align}
Notice that the fusion rule \eqref{fusionruleR2=N} implies that indeed $(\text{dim}\,\cT)^2=N$.

Using \eqref{TV} as a starting point, we now discuss the action of the $\cT$-defect on the physical spectrum of closed string theory. We consider, in turn, massless and massive states.

\subsection{Action of $\cT$ symmetry on closed string states}
Let us now see how this symmetry acts on the massless vertex operators in the low-energy theory. Upon compactifying the 25$^{\rm th}$ direction, the metric, the $B$-field, and the dilaton split as
\begin{align}
    G_{IJ}&\rightarrow (G_{\mu\nu}, A_\mu, e^{2\sigma})\,, & B_{IJ}& \rightarrow (B_{\mu\nu}, C_\mu)\,, &\Phi_D\rightarrow\Phi_d + \s \,,
\end{align}
where $A_\mu\equiv G_{\mu \, 25}$, $e^{2\s}\equiv G_{25 \, 25}$, and $C_{\mu} \equiv B_{\mu \,25}$. We thus have the following massless vertex operators:
\begin{itemize}
    \item $G_{\mu\nu}$, $B_{\mu\nu}$, and $\Phi_d$ correspond respectively to the symmetric traceless, antisymmetric, and trace parts of the vertex operator $V^{\mu\nu} = \half g^{z\Bar{z}}\normOrd{\partial X^\mu \, \Bar{\partial}X^\nu e^{ik\vdot X}}$
    \item $A_\mu$ and $C_\mu$ correspond respectively to $V_\pm^{\mu}=\half g^{z\Bar{z}}\normOrd{\,(\partial X^\mu \, \Bar{\partial}\phi\pm \partial \phi \, {\partial}X^{\mu}) e^{ik\vdot X}}$
    \item $\sigma$ corresponds to the vertex operator $V_\sigma=\half R^2
g^{z\Bar{z}}\normOrd{\partial \phi\,\Bar{\partial}\phi\,e^{ik\vdot X}}$
\end{itemize}
where $\normOrd{\,}$ denotes normal ordering, $k\vdot X=k_\mu X^\mu$ with $k_\mu$ being the momentum along the non-compact directions, $R^2$ is the background value of $e^{2\sigma}$, and we adopted complex coordinates on the worldsheet (see appendix \ref{conventions} for details).
Hence, we see that the $\cT$-defect acts on the massless vertex operators as
\begin{align}\label{TVsigma}
  \cT \ket{V^{\mu\nu}} = \sqrt{N}\ket{V^{\mu\nu}}\,, \qquad   \cT \ket{V_\sigma} = - \sqrt{N}\ket{V_\sigma}\,, \qquad  \cT \ket{V^\mu_\pm} = - \sqrt{N} \ket{V^\mu_\mp}~.
\end{align}

Let us now look at the action of the $\cT$-defect on generic massive vertex operators, which are allowed to carry momentum and winding charges. We consider
\begin{align}\label{Vmw}
    \mathsf{V}_{m,w}(z,\Bar{z}) = \normOrd{\,\exp\left(im\phi(z,\Bar{z})+iw\Tilde{\phi}(z,\Bar{z})\right)\,}~,
\end{align} 
which carries charges $(m,w)$ under the $\U(1)_m\times\U(1)_w$ symmetry. Notice that any of the massless vertex operators can be dressed with $\mathsf{V}_{m,w}$, thus defining operators carrying non-trivial charges.

The action of the $\cT$-defect on the state $\ket{m,w}$ corresponding to $\mathsf{V}_{m,w}$ can be deduced using \eqref{1dtr} and it reads
 \begin{align}\label{Tact}
     \cT \ket{m,w}  =  \begin{cases}
        \sqrt{N}e^{i\pi m w}\,  \ket{Nw,\dfrac{m}{N}} &\text{ when }N\;|\;m\,,\\
        0 &\text{ when }N\nmid m\,.
    \end{cases}
 \end{align}
The phase in \eqref{Tact} is determined by the action of T-duality \cite{Fuchs:2007tx}, and it is consistent with the general case analyzed in appendix \ref{app:phase}. Notice that when $R^2 \neq 1$ the $\cT$ symmetry has a non-trivial kernel, and hence it is non-invertible (equivalently, $M_B \in \tO(1,1;\integers)$ if and only if $N=1$).

\begin{figure}[t]
    \centering
    \begin{center}
    \begin{tikzpicture}[scale=1]
        \node[label={-45:$\cT$}] at (0.5,0) {};
        \fill[fill=black] (-1.1,2.1) circle (1pt) node[anchor = north west]{$\mathsf{V}_{m,w}$};
        \draw[thick]{} (1,0)--(1,4);
        \node[label={-45:$=\quad e^{i\pi m w }$}] at (2.51,2.5) {};
        \fill[fill=black] (8.1,2) circle (1pt) node[anchor = north west]{$\mathsf{V}_{Nw,\frac{m}{N}}$};
        \draw[dashed,]{} (8,2)--(6,2);
        \node[label={-45:$\eta^{m}$}] at (6.6,3) {};
        \draw[thick] (6,0)--(6,4);
        \node[label={-45:$\cT$}] at (5.5,0) {};
        \draw[<-]  (7,2)--(7.1,2);
\end{tikzpicture} \end{center}
\caption{Action of the topological defect $\cT$ on the vertex operator $\mathsf{V}_{m,w}$. If $m$ is not a multiple of $N$, the resulting vertex operator is non-genuine and needs to be placed at the end-point of the momentum $\integers_N$ defect $\eta^m$ for gauge invariance.}
\label{fig:defectOp}
\end{figure}
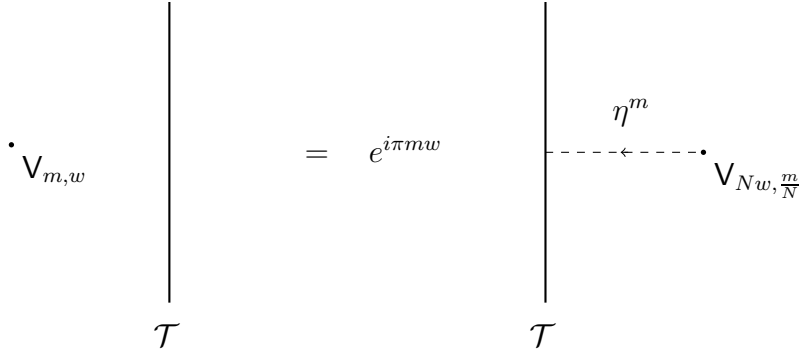

As already mentioned at the beginning of this section, we can describe the action of the $\cT$-defect in the language of operators rather than states, by dragging it across an insertion of $\mathsf{V}_{m,w}$ at the point $(z,\Bar{z})$, as depicted in figure \ref{fig:defectOp}. In correspondence with the two cases in \eqref{Tact}, we have the following two possibilities.

\begin{itemize}
    \item Consider a vertex operator $\mathsf{V}_{m,w}$ with $N\,|\,m$. It is clear from \eqref{Tact} that the result of the defect action is the genuine vertex operator $\mathsf{V}_{Nw,m/N}$, which has integer momentum and winding charges. Consistently, the momentum line in figure \ref{fig:defectOp} trivializes, as $\eta^N=1$. 

    \item Consider a vertex operator $\mathsf{V}_{m,w}$ with $N\nmid m$. When we drag the defect across it, we end up with a non-genuine vertex operator that needs to be placed at the end-point of a momentum defect $\eta^u$ because of gauge invariance. To identify the value of $u$, we use the first expression in \eqref{etadefinition} to require that
    \begin{align}
    \label{endpoint}
    \normOrd{\,e^{-i \frac{u}{N} \Tilde{\phi}} \, \mathsf{V}_{Nw, \frac{w}{N}}\,} = \normOrd{\,e^{iNw\,\phi+i\frac{m - u}{N}\Tilde{\phi}}\,}
    \end{align}
    is gauge-invariant under periodic identifications of the dual scalar  $\Tilde{\phi} \sim \Tilde{\phi}+2\pi$. This imposes that $u=m$ mod $N$ (the negative sign $-u$ has been chosen for consistency with the orientation of the $\eta$ line shown in figure \ref{fig:defectOp}), i.e.~$\eta^u=\eta^m$.\\
\end{itemize}
\subsection{Selection rules for the non-invertible $\cT$ symmetry}
In this subsection, we analyze the selection rules for the $\cT$ symmetry on the sphere $\amsmathbb{S}^2$ and on the torus $\amsmathbb{T}^2$.
For concreteness, we focus on the one-point function of the vertex operator $V_\sigma$ in the presence of a $\cT$-defect.
Since the spacetime momentum $k_\mu$ plays no role in our discussion, we simply consider an insertion with $k=0$, namely
\begin{align}
V_\sigma(z,\Bar{z}) = \half  R^2 g^{z\Bar{z}}\normOrd{\,\partial \phi(z)\Bar{\partial} \phi(\Bar{z})\,} \,.
\end{align}
While $\expval{V_\sigma}=0$ on $\amsmathbb{S}^2$, as required by the selection rules on the sphere, we show explicitly that the non-vanishing of $\expval{V_\sigma}$ on $\amsmathbb{T}^2$, which can be computed explicitly, is compatible with the selection rules for the non-invertible $\cT$ symmetry on the torus (and, in principle, on a worldsheet of higher genus).

\begin{figure}[h]
\centering
\begin{tikzpicture}[scale = .6, baseline = 0]
  \shade[ball color = gray!40, opacity = 0.4] (0,0) circle (3cm);
  \draw (0,0) circle (3cm);
  \draw (-3,0) arc (180:360:3 and 0.6);
  \draw[dashed] (3,0) arc (0:180:3 and 0.6);
  \draw[] (-1,1.7) circle (0.85cm);
  \node[] at (0.1,2.5) {$\cT$};
  \fill[fill=black] (-1,1.8) circle (1pt) node[below]{$V_\s$};
\end{tikzpicture}
$=$
\begin{tikzpicture}[scale = .6, baseline = 0]
  \shade[ball color = gray!40, opacity = 0.4] (0,0) circle (3cm);
  \draw (0,0) circle (3cm);
  \draw (-3,0) arc (180:360:3 and 0.6);
  \draw[dashed] (3,0) arc (0:180:3 and 0.6);
  \draw[] (1,-1.8) circle (0.85cm);
  \node[] at (-0.1,-2.5) {$\cT$};
  \fill[fill=black] (-1,1.8) circle (1pt) node[below]{$V_\s$};
\end{tikzpicture}
\caption{Selection rule on $\amsmathbb{S}^2$. The equality follows by dragging the $\cT$-defect around the sphere, without crossing the insertion of $V_\sigma$.}
\label{fig:WS2}
\end{figure}
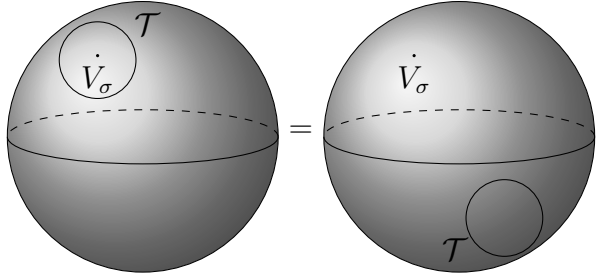
On the sphere $\amsmathbb{S}^2$, we can easily obtain the selection rule as in figure \ref{fig:WS2}. On the left, we can act on $V_\sigma$ with the $\cT$-defect which we then shrink, producing a factor of the quantum dimension, as in \eqref{TVsigma}. On the right, we can simply shrink the $\cT$-defect without crossing the insertion. We thus get
\begin{align}\label{WardS2}
-\dim(\cT)\expval{V_\sigma(z,\Bar{z})}_{\amsmathbb{S}^2} = \dim(\cT)\expval{V_\sigma(z,\Bar{z})}_{\amsmathbb{S}^2}\quad\Rightarrow\quad\expval{V_\sigma(z,\Bar{z})}_{\amsmathbb{S}^2} = 0~.
\end{align}
Using Wick's theorem, it is straightforward to verify the above selection rule.

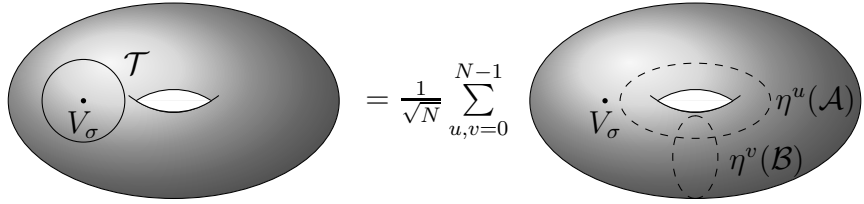
\begin{figure}[h]
\centering
    \begin{tikzpicture}
        \draw[ball color = gray!40, opacity = 0.4] (1,0) ellipse (2.2cm and 1.3cm);
        \draw[smooth] (0.4,0.07) .. controls (0.8,-0.23) and (1.2,-0.23) .. (1.6,0.07);
        \draw[smooth,fill = white] (0.5,0) .. controls (0.8,0.2) and (1.2,0.2) .. (1.5,0); 
        \draw[smooth,fill = white] (0.5,0) .. controls (0.8,-0.2) and (1.2,-0.2) .. (1.5,0); 
        \draw[] (-0.2,0) circle (0.55cm);
        \fill[fill=black] (-0.2,0) circle (1pt) node[below]{$V_\s$};
        \node at (4.5,0) {$= \frac{1}{\sqrt{N}}\sum\limits_{u,v=0}^{N-1}$};
        \node[] at (0.5,0.5) {$\cT$};
    \end{tikzpicture}
    \begin{tikzpicture}
        \draw[ball color = gray!40, opacity = 0.4] (1,0) ellipse (2.2cm and 1.3cm);
         \draw[smooth] (0.4,0.07) .. controls (0.8,-0.23) and (1.2,-0.23) .. (1.6,0.07);
        \draw[smooth,fill = white] (0.5,0) .. controls (0.8,0.2) and (1.2,0.2) .. (1.5,0);  
        \draw[smooth,fill = white] (0.5,0) .. controls (0.8,-0.2) and (1.2,-0.2) .. (1.5,0); 
        \node[dashed,ellipse,
        		draw = black,
        		text = violet,
        		minimum width =2 cm, 
        		minimum height = 1cm] (e) at (1,0) {};		
        \draw[dashed,color = black] (1,-1.3) arc(270:85:0.3 and 0.55);
        \draw[dashed,color = black] (1,-1.3) arc(270:450:0.3 and 0.55);
        \fill[fill=black] (-0.2,0) circle (1pt) node[below]{$V_\s$};
        \node[] at (2.6,0) {$\eta^u(\cA)$};
        \node[] at (1.95,-0.8) {$\eta^v(\cB)$};
    \end{tikzpicture}
    \caption{Selection rule on $\amsmathbb{T}^2$ for $R^2 = N$. The equality follows by dragging the $\cT$-defect without crossing the insertion of $V_\sigma$, and applying the fusion rule of the $\cT$-defect and the F-symbols.}
    \label{fig:WT2}
\end{figure}

On the torus $\amsmathbb{T}^2$, the selection rule for $R^2=N$ is shown in figure \ref{fig:WT2}.
One can easily derive it by using the fusion rule of the $\cT$-defect in \eqref{fusionruleR2=N} and the F-symbols (see e.g.~eq.~$(4.9)$ of \cite{Chang:2018iay}). On the left, we can act on $V_\sigma$ with the $\cT$-defect which we then shrink, so that the selection rule on $\amsmathbb{T}^2$ takes the form
\begin{align}
    \expval{ V_\sigma(z,\Bar{z})}_{\amsmathbb{T}^2} =- \frac1N  \sum_{\gamma \in H_1(\amsmathbb{T}^2, \mathbbm{Z}_N)}\expval{\eta(\gamma)V_\sigma(z,\Bar{z})}_{\amsmathbb{T}^2}~.
\end{align}

\noindent It is convenient to rewrite this explicitly as
\begin{align}\label{WardT2}
    \expval{ V_\sigma(z,\Bar{z})}_{\amsmathbb{T}^2} =- \frac1N \sum_{u,v=0}^{N-1}\expval{\eta^u(\cA)\eta^v(\cB)V_\sigma(z,\Bar{z})}_{\amsmathbb{T}^2}~,
\end{align}
where $\{\cA, \cB\}$ is a basis for $H_1(\amsmathbb{T}^2, \mathbbm{Z}_N) =\integers_N\times \integers_N$ and $u,v\in\integers_N$. Notice that, as opposed to the case of the sphere, here the selection rule does not imply the vanishing of $\expval{V_\sigma(z,\Bar{z})}$ on the torus.  On the torus, the operator $V_\sigma$ is
\begin{align}\label{VsigT2}
    V_\sigma(z,\Bar{z}) = \tau_2 R^2 \normOrd{\,\partial \phi(z)\Bar{\partial}\phi(\Bar{z})\,}~,
\end{align}
where $\tau = \tau_1+i\tau_2$ (with $\tau_2 >0$) is the torus modular parameter (see appendix \ref{conventions} for a summary of our conventions).

We now verify explicitly that the selection rule is satisfied by calculating each term that appears in \eqref{WardT2}. To do so, we evaluate the path integral by \textit{chiral splitting}, closely following \cite{DHoker:2022dxx,Aoki:2004sm}.
To compute the terms on the right-hand side, notice that the insertion of the momentum defects $\eta^u(\cA)$ and $\eta^v(\cB)$ corresponds, respectively, to giving $\phi$ monodromies around the non-trivial $\cA$ and $\cB$ cycles of the torus.
Indeed, the action of the $\eta$ defect (generating $\integers_{N}\subset \U(1)_m$) on the state $\ket{m,w}$ corresponding to the vertex operator $\mathsf{V}_{m,w}$ is
\begin{align}
\eta\ket{m,w} = e^{\frac{2\pi i m }{N}} \ket{m,w} \,.
\end{align}
It then follows that in presence of the insertions $\eta^u(\cA)$ and $\eta^v(\cB)$, the vertex operator $\mathsf{V}_{m,w} (z,\Bar{z})$, defined in \eqref{Vmw}, has monodromies
\begin{align}\label{Vmon}
    \mathsf{V}_{m,w}(z+\tau)=e^{\frac{2\pi i um}{N}}\;\mathsf{V}_{m,w}(z)\,,\qquad \mathsf{V}_{m,w}(z+1)=e^{\frac{2\pi i vm}{N}}\;\mathsf{V}_{m,w}(z)~,
\end{align}
around the $\cA$ and $\cB$ cycles, respectively. These correspond to $\phi$ having monodromies
\begin{align}\label{monodromies}
    \phi(z+\tau) &= \phi(z) + 2\pi  \left(m+ \frac{u}{N}\right) \,, & \phi(z+1) &= \phi(z) + 2\pi \left(w+\frac{v}{N}\right) \,.
\end{align}
Hence, restricting to each $(u,v)$ sector characterized by these monodromies correspond to inserting $\eta^u(\cA)$ and $\eta^v(\cB)$ lines in the one-point function.
Below we use this fact to evaluate each term on the right-hand side of \eqref{WardT2} using the functional integral.

First, we define a field configuration $\phi_{m,w}$, satisfying \eqref{monodromies}, as
\begin{align}\label{WSinst}
    \phi_{m,w}(z,\Bar{z}) = \varphi(z,\Bar{z}) + 2\pi  z \frac{\left(m+\frac{u}{N}\right)-\Bar{\tau}\left(w+\frac{v}{N}\right)}{\tau-\Bar{\tau}}-2\pi  \Bar{z} \frac{(m+\frac{u}{N})-{\tau}(w+\frac{v}{N})}{\tau-\Bar{\tau}}~,
\end{align}
where $\varphi(z) = \varphi(z+1) = \varphi(z+\tau)$ is periodic around $\cA$ and $\cB$, whereas the other terms capture the monodromies.
The field configuration in \eqref{WSinst} is called a \textit{worldsheet instanton} and its action \eqref{eq:compact_action} reads (in the $d=1$ case, where $G=R^2$ and $B=0$)
\begin{align}
    S[\phi_{m,w}] = S[\varphi] + \frac{\pi R^2}{\tau_2} \abs{\left(m+\frac{u}{N}\right)-\tau \left(w+\frac{v}{N}\right)}^2~.
\end{align}
Plugging the parametrization \eqref{WSinst} into the expression \eqref{VsigT2} for $V_\s$, we obtain
\begin{align}\label{twoTerms}
     V_\sigma  = \tau_2R^2 {\partial \varphi \, \Bar{\partial}\varphi} + \frac{\pi^2 R^2}{\tau_2}\abs{\left(m+\frac{u}{N}\right) - \tau\left(w+\frac{v}{N}\right)}^2 + \dots\,,
\end{align}
where the dots denote terms that are linear in $\partial \varphi$ or $\Bar{\partial}\varphi$, which we discard since they vanish upon functional integration over $\varphi$.

We now evaluate the contribution of both terms in \eqref{twoTerms} to the functional integral. The parametrization \eqref{WSinst} is useful because for any functional $\cF$ on $\amsmathbb{T}^2$ we have
\begin{align}
    \expval{\cF[\phi]} = \frac{1}{Z} \int D\phi \, \cF[\phi] e^{-S[\phi]} = \frac{R}{Z} \sum_{m,w\in\integers}\int D\varphi \, \cF[\phi_{m,n}]e^{-S[\phi_{m,n}]}\,,
\end{align}
where the factor of $R$ comes from the zero-mode of the Laplacian on $\amsmathbb{T}^2$, and the measure $D\varphi$ refers to the space of orthogonal modes (on which the propagator can be defined).
There are two terms in $\expval{\eta^u(\cA)\eta^v(\cB) V_\s}$ that need to be evaluated, corresponding to the two terms in \eqref{twoTerms}.

\begin{itemize}
    \item The first contribution to $\expval{\eta^u(\cA)\eta^v(\cB) V_\s}$ is
    \begin{align}
    \tau_2R^2\expval{\eta^u(\cA)\eta^v(\cB)\partial\varphi\,\Bar{\partial}\varphi}
    &= \tau_2R^2 \left(\frac{1}{Z}\int D\varphi \, {\partial \varphi\,\Bar{\partial}\varphi} \, e^{-S[\varphi]}\right)Z_{u,v} \,,
    \label{varphiPI}
    \end{align}
    where $Z_{u,v}$ is the partition function with the $\eta^u$ line inserted along the $\cA$ cycle and the $\eta^v$ line inserted along the $\cB$ cycle, namely
    \begin{align}
    Z_{u,v} &= \frac{R}{\abs{\eta(\tau)}^2}\sum_{m,w\in\integers}e^{-\frac{\pi R^2}{\tau_2} \abs{\left(m+\frac{u}{N}\right)-\tau \left(w+\frac{v}{N}\right)}^2} \,,
    \end{align}
    and clearly $Z_{0,0}=Z$. The factor in parentheses of \eqref{varphiPI} can be evaluated by point-splitting, and using the well-known propagator of $\varphi$ on the torus (see e.g.~\cite{DHoker:2022dxx}), which in our conventions reads
    \begin{align}\label{shift}
    \expval{\varphi(z)\varphi(w)} = - \frac{1}{2 R^2} \log\abs{\frac{\vartheta_1(z-w|\tau)}{\eta(\tau)}}^2 + \frac{\pi}{\tau_2 R^2}[\Im (z-w)]^2~,
    \end{align}
    where $\eta(\tau)$ is the Dedekind $\eta$-function and $\vartheta_1(z|\tau)$ is the first Jacobi $\vartheta$-function. We thus get
    \begin{align}
    \tau_2R^2\expval{\eta^u(\cA)\eta^v(\cB)\partial\varphi\,\Bar{\partial}\varphi} & =
    - \tau_2 R^2 \frac{Z_{u,v}}{Z} \lim_{w\rightarrow z} \partial\Bar{\partial}\expval{\varphi(z)\varphi(w)} = - \frac{\pi Z_{u,v}}{2 Z} \,,
    \end{align}
    where we have subtracted the constant divergence proportional to $\delta(0)$ arising from taking the limit of coincident points, according to the normal-ordering prescription.
    
    \item The second contribution to $\expval{\eta^u(\cA)\eta^v(\cB) V_\s}$ is
    \begin{align}
   \frac{R}{Z\abs{\eta(\tau)}^2} \sum_{m,w\in\integers}  \frac{\pi^2R^2 }{\tau_2} \abs{(m+u)-\tau (w+v)}^2 e^{-\frac{\pi R^2}{\tau_2} \abs{(m+u)-\tau (w+v)}^2} &= \frac{\pi Z_{u,v}}{2Z} - \frac{\pi R}{2Z} \frac{\partial Z_{u,v}}{\partial R} \,.
    \end{align}
    
\end{itemize}
Combining both contributions, we obtain
\begin{align}\label{WardT2RHS}
\expval{\eta^u(\cA)\eta^v(\cB) V_\s} = -\frac{\pi R}{2 Z}\frac{\partial Z_{u,v}}{\partial R}\,.
\end{align}
Note that this relation should have been expected since $V_\sigma$ is proportional to the Lagrangian,%
\footnote{Indeed, writing the action in complex coordinates, following our conventions of appendix \ref{conventions}, the path integral for each $(u,v)$ sector is
\begin{align}
    Z_{u,v} &= \int D\phi\; \eta^u(\cA)\eta^v(\cB)\exp\left(-\frac{R^2}{2\pi} \int_{\amsmathbb{T}^2} d^2 z \, \partial \phi \, \Bar{\partial}\phi\right) \,.
\end{align}
The one-point function of $V_\sigma(z,\Bar{z})$ should be independent of the position $z$, so we can get such correlator by deriving the path integral with respect to $R^2$ as
\begin{align}\label{KeyEqn}
    \expval{\eta^u(\cA)\eta^v(\cB)V_\sigma} &= \frac{\expval{\int_{\amsmathbb{T}^2} d^2 z\;\eta^u(\cA)\eta^v(\cB) V_\s}}{2\tau_2}= -\frac{\pi R}{2Z} \frac{\partial Z_{u,v}}{\partial R} \,,
\end{align}
where $\int_{\amsmathbb{T}^2} d^2 z = 2 \tau_2$ is the appropriate volume normalization factor.}
but the method presented here is general and can be applied to any operator.

In appendix \ref{T2Ward_app2}, we explicitly evaluate $Z_{u,v}$ and verify that the selection rule \eqref{WardT2} is satisfied. A quick way to see that this is indeed the case is to use \eqref{WardT2RHS} to rewrite the selection rule as
\begin{align}
    R \frac{\partial Z(R)}{\partial R}\bigg|_{R=\sqrt{N}} = - \frac{1}{N} \sum_{u,v=0}^{N-1} R \frac{\partial Z_{u,v}(R)}{\partial R}\bigg|_{R=\sqrt{N}} \,.
\end{align}
On the right-hand side, we notice that summing over insertions of $\eta$ lines along non-trivial cycles is equivalent to gauging the discrete $\integers_N$ subgroup of $\U(1)_m$, whose net effect is to rescale the radius as $R\rightarrow R/N$. We are then left to verify that
\begin{align}\label{wardwithZ}
    R \frac{\partial Z(R)}{\partial R}\bigg|_{R=\sqrt{N}} = - R \frac{\partial Z(R/N)}{\partial R}\bigg|_{R=\sqrt{N}} \,.
\end{align}
Using on the right-hand side that the partition function is invariant under T-duality, namely $Z(R/N)=Z(N/R)$, shows that the selection rule is satisfied. Notice that \eqref{wardwithZ} also implies $Z'(1)=0$, so that $\expval{V_\s}_{\amsmathbb{T}^2}=0$ at $R=1$. This is consistent with the fact that at the self-dual point $R=1$ the $\cT$-defect corresponds to an invertible $\integers_2$ symmetry, under which $V_\sigma$ is odd. In this case, the selection rules imply the vanishing of such correlator on surfaces with arbitrary genus.

\subsection{Comments on string perturbation theory}
As we have shown in the previous section, selection rules are typically modified on a Riemann surface with non-trivial topology. For generic non-invertible symmetries this is clear from the fact that the derivation of selection rules on a higher-genus worldsheet involves a non-trivial network of topological lines. Ward identities are also known to get modified on higher-genus surfaces, an example is the conformal Ward identity \cite{Eguchi:1986sb}.

The selection rules for the $\cT$-defect at $R^2=N$ on a Riemann surface $\Sigma_g$ of genus $g$ can be obtained by generalizing the arguments of the previous section as
\begin{equation}\label{SigmagWard}
   \left< \cT \cdot V_\s \right>_{R,\Sigma_g} = \frac{1}{\sqrt{N}}\sum_{\g\in H_1(\Sigma_g, \amsmathbb{Z}_N)}\left< \eta(\g) V_\s \right>_{R,\Sigma_g} \,.
\end{equation}
In string perturbation theory, amplitudes are defined by
\begin{equation}
    \langle  \langle V_\s \rangle  \rangle_R = \sum\limits_{g=0}^\infty g_s^{2g-2} \int_{\mathcal M_g} \left< \cT \cdot V_\s \right>_{R,\Sigma_g} \,,
\end{equation}
where $\cM_g$ is the moduli space of the surface $\Sigma_g$.
Hence, we have 
\begin{align}\label{stringamplitude}
     \langle  \langle \cT \cdot V_\s \rangle  \rangle_R & = \sum\limits_{g=0}^\infty g_s^{2g-2}\frac{1}{\sqrt{N}} \int_{\mathcal M_g} \sum_{\g\in H_1(\Sigma_g, \amsmathbb{Z}_N)} \left< \eta(\g) V_\s \right>_{R,\Sigma_g} \no\\
     & = \sqrt{N}\langle  \langle V_\s \rangle  \rangle_{1/R} \,,
\end{align}
where we have used the fact that summing over all insertions of $\mathbbm{Z}_N$ lines is equivalent to gauging, namely $R \rightarrow R/N = 1/R$.
Since the action of the $\cT$-defect is $\cT \cdot V_\s = - \sqrt{N} \, V_\s$, the relation \eqref{stringamplitude} is precisely the statement of T-duality
\begin{equation}
    \langle  \langle  V_\s \rangle  \rangle_R  =-\langle  \langle V_\s \rangle  \rangle_{1/R}~.
\end{equation}
In other words, when considering correlators of local operators, the selection rules for the non-invertible defects constructed by combining dualities and gauging are satisfied in string perturbation theory as a consequence of the corresponding duality element (T-duality, in the simple case analyzed here) of the worldsheet theory.

The situation is more interesting when we consider correlation functions involving non-local operators on a higher-genus surface. As an example, consider the two-point function of two vertex operators of opposite charges $\mathsf{V}_{m,w}$ and $\mathsf{V}_{-m,-w}$. Applying the arguments of the previous section on the torus, we get (omitting for simplicity the dependence on the coordinates)
\be
\langle \mathsf{V}_{m,w} \mathsf{V}_{-m,-w} \rangle_{1/R} =
\langle \mathsf{V}_{Nw,\frac{m}{N}} \, \eta^m \, \mathsf{V}_{-Nw,-\frac{m}{N}} \rangle_{R} \,,
\ee
where $\eta$ is the topological $\integers_N$ defect line defined in \eqref{etadefinition} that is necessary for gauge invariance if $m \neq 0$ mod $N$. This relation involves a correlator with an insertion of a topological line and it gives a selection rule which is a genuine consequence of the non-invertible symmetry. As in the previous example, this can be extended to amplitudes on a arbitrary worldsheet $\Sigma_g$.

Notice that the discussion of this section is valid only in the framework of string perturbation theory, whereas in the full non-perturbative setup we do not expect global symmetries, including non-invertible ones \cite{Heckman:2024obe}.

\section{Outlook}

In this work, we constructed topological defects in a two-dimensional worldsheet theory, arising from bosonic string theory compactified on a torus. We conclude with a few interesting open questions. 

It would be interesting to analyze other compactifications of bosonic string theory. It is known that the worldsheet theory admits dualities in other compactifications as well (see e.g.~\cite{Giveon:1994fu}). Moreover, one could try to extend the discussion to superstring theory, possibly considering non-zero RR-fluxes (for instance, see \cite{Elitzur:2013ut} for a discussion on topological interfaces acting as T-duality).
It would also be interesting if one can understand the $\SL(2,\mathbbm{Z})$ duality of the target space low-energy supergravity theory in $AdS_5\times S^5$ as a topological defect on the worldsheet. 

In this work, we constructed topological defects using the half-space gauging approach. We closely followed \cite{Niro:2022ctq}, where an infinite set of non-invertible defects, acting on fields as $\SO(2)$ rotations and on charges as $\SL(2,\rationals)$ transformations, was constructed in 4d Maxwell theory. Similarly, we assumed that certain rationality conditions are satisfied, and in particular in section \ref{section: Topological defects on the worldsheet} we realized defects acting on fields as $\tO(d)\times\tO(d)$ rotations and on charges as elements of $\tO(d,d;\rationals)$. In the Lagrangian construction of section \ref{section: Lagrangian description of the topological defects}, these rationality conditions were necessary for the gauge invariance of the defect Lagrangian. It has been recently proposed \cite{Hasan:2024aow} that one can realize in 4d Maxwell theory the whole $\SO(2)$ action. It would be interesting to understand whether one can use similar arguments to relax our rationality conditions and construct non-invertible defects corresponding to arbitrary elements of $\tO(d)\times\tO(d)$.

A complementary viewpoint is provided by the SymTFT approach for continuous symmetries \cite{Brennan:2024fgj,Antinucci:2024zjp,Bonetti:2024cjk,Apruzzi:2024htg}, which has been used to construct non-invertible axial rotations of any angle in 4d massless QED \cite{Arbalestrier:2024oqg} and the $\cT$-defect for $d=1$ at any (positive) real value of $R^2$ \cite{Arguriotoappear}. It would be interesting to understand the (non-Abelian) SymTFT description of the defects studied in our work for generic $d>1$, and possibly construct the full $\tO(d)\times\tO(d)$ action.

%
%
%
%
%
\section*{Acknowledgements}
We would like to thank Riccardo Argurio, Eric D'Hoker, Michael Gutperle, Yan-Yan Li, Lukas Lindwasser, Jonathan Heckman, and Orr Sela for valuable discussions. SB and PN are supported by the Mani L. Bhaumik Institute for Theoretical Physics. PN is also supported by a DOE Early Career Award under DE-SC0020421. KR is supported by the Simons Collaboration on Global Categorical Symmetries and also by the NSF grant PHY-2112699.
\appendix

\section{Conventions}\label{conventions}
In this appendix, we summarize our conventions for coordinates on a two-dimensional Euclidean worldsheet $\Sigma$. We consider in turn the cases of $\Sigma = \amsmathbb{S}^2$ and $\Sigma =\amsmathbb{T}^2$.

On the sphere $\amsmathbb{S}^2$, we pick local coordinates $z = \sigma^1+i\sigma^2$ and $\Bar{z} = \sigma^1 - i\sigma^2$. The metric and the $\ep$ tensor read (we take $\ep_{12} = +1$)
\begin{align}
    g_{z\Bar{z}} &= \half \,, &g_{zz} &= g_{\Bar{z}\Bar{z}} = 0  \,,
    &\ep_{z\bar{z}} &= \frac{i}{2} \,, &\ep^{z\bar{z}} &= -2i \,.
\end{align}
In these coordinates, the integration measure is
\begin{align}
   \text{Vol}_{\amsmathbb{S}^2} = \sqrt{g}\,d\sigma^1 \wedge d\sigma^2=\sqrt{g}\, idz\wedge d\Bar{z}\equiv \half d^2z \,,
\end{align}
and the differential operators are
\begin{align}
    \partial &= \frac{\partial_1 - i\partial_2}{2} \,, &\Bar{\partial} &= \frac{\partial_1+i\partial_2}{2} \,.
\end{align}
Notice that this is the set of coordinates we use in sections \ref{section: Topological defects on the worldsheet} and \ref{section: Lagrangian description of the topological defects} on a worldsheet $\Sigma$ with generic topology.
However, in the case of the torus it is sometimes convenient to choose a different set of coordinates, that we use in section \ref{section: defects at d=1}. On a torus $\amsmathbb{T}^2$ with modulus $\tau = \tau_1+i\tau_2$ (with $\tau_2>0$), we pick local coordinates $z = \sigma^1+\tau \sigma^2$ and $\Bar{z} = \sigma^1 +\Bar{\tau} \sigma^2$, where $\sigma^1$ and $\sigma^2$ have unit periods. The metric and the integration measure are
\begin{align}
    g_{z \Bar{z}} &= \frac{1}{2\tau_2} \,, \quad &g_{zz} &= g_{\Bar{z}\Bar{z}} = 0  \,, \quad & \text{Vol}_{\amsmathbb{T}^2} = \sqrt{g}\,\,d\sigma^1 \wedge d\sigma^2 &=\frac{i}{2\tau_2}dz\wedge d\Bar{z}\equiv \frac{1}{2\tau_2} d^2 z \,.
\end{align}
In these coordinates, the differential operators are
\begin{align}
    \partial &= -\frac{i}{2\tau_2}(\partial_2 - \Bar{\tau}\partial_1) \,,  &\Bar{\partial} = \frac{i}{2\tau_2}(\partial_2 - \tau\partial_1) \,.
\end{align}


\section{Phase in the duality action on vertex operators}
\label{app:phase}

In this appendix, we determine the phase in the action of an $\tO(d,d;\integers)$ duality transformation on vertex operators. This phase appears in the action on states \eqref{eq:actionStates} of each duality component $\cM$ of non-invertible defects. The operator product expansion of two vertex operators with charges $\mathbf{q}_1$ and $\mathbf{q}_2$ is\footnote{We are implicitly dressing the vertex operators with the so-called \textit{cocycle} factors, which guarantee the validity of bosonic commutation relations.}

\begin{equation}
    V_{\mathbf{q}_1}(z,\Bar{z}) V_{\mathbf{q}_2}(w,\Bar{w}) \sim (-)^{ \mathbf{q}_1^T Q \, \mathbf{q}_2} \, (z-w)^{\frac{1}{2}\mathbf{q}_1^T Z_+ \, \mathbf{q}_2} \, (\Bar{z}-\Bar{w})^{\frac{1}{2}\mathbf{q}_1^T Z_- \, \mathbf{q}_2} \, V_{\mathbf{q}_1 +\mathbf{q}_2}(w,\Bar{w}) ~,
\end{equation}
where the non-trivial phase on the right-hand side is necessary for locality (see e.g.~\cite{Fuchs:2007tx}), and
\begin{equation}
    Q=
    \begin{pmatrix}
        0 & \mathbbm 1_d \\
        0 & 0
    \end{pmatrix}~, \qquad
    Z_\pm = Z(G,B) \pm J \,.
\end{equation}
The dependence on the coordinates does not play any role in the following, as the combination $\mathbf{q}^T Z_\pm \mathbf{q}$ is obviously invariant under duality transformations.
Let us consider a duality element $\cM\in \tO(d,d;\mathbbm Z)$ acting as
\begin{equation}
    V_{\mathbf{q}} \rightarrow e^{i \theta_\cM(\mathbf{q})} V_{\cM\mathbf{q}}~.
\end{equation}
Consistency with the operator product expansion of vertex operators requires 
\begin{equation}
\label{constraint_on_theta}
    \theta_\cM(\mathbf{q}_1+\mathbf{q}_2)-\theta_\cM(\mathbf{q}_1)-\theta_\cM(\mathbf{q}_2) = \pi\; \mathbf{q}_1^T (Q - \cM^T Q \cM) \mathbf{q}_2 \quad \text{mod }2\pi~.
\end{equation}
Notice that in the case of a T-duality transformation performed simultaneously on all fields we have that $\cM=J$, and the phase is precisely given by the Dirac pairing
\begin{equation}
\label{theta_for_tduality}
\theta_J(\mathbf{q}) = \frac{\pi}{2} \mathbf{q}^T J \mathbf{q} \quad \text{mod }2\pi\,,     
\end{equation}
which indeed solves \eqref{constraint_on_theta}. For $d=1$, this gives the phase in \eqref{Tact}.

For generic $\cM$, we can view \eqref{constraint_on_theta} as a recursion relation.
Using that $Q-\cM^T Q \cM$ is antisymmetric for any $\cM\in\tO(d,d;\integers)$, the solution is (up to an arbitrary linear term in $\mathbf{q}$ which can be set to zero using momentum and winding transformations)
\begin{equation}
    \theta_\cM(\mathbf{q}) = \pi \sum_{a>b}^{2d}\mathbf{q}_a (Q- \cM^T Q \cM)_{ab} \mathbf{q}_b \quad \text{mod }2\pi~.
\end{equation}
It is easy to verify that in the case of those $\widetilde{\cM} \in \tO(d,d;\integers)$ such that
\begin{equation}
\widetilde{\cM}^T Q \widetilde{\cM}=Q^T \quad \Leftrightarrow \quad \widetilde{\cM} = \begin{pmatrix}
0 & \beta \\ \beta^{-T} & 0    
\end{pmatrix} \, \text{ with } \beta \in \text{GL}(d;\integers) \,,  
\end{equation}
the solution $\theta_{\widetilde\cM}(\mathbf{q})$ can be rewritten as the bilinear form on the right-hand side of \eqref{theta_for_tduality}.


\section{Verifying the selection rule for the $\cT$ symmetry on $\amsmathbb{T}^2$}
\label{T2Ward_app2}

In this appendix, we explicitly compute each term on both sides of the selection rule \eqref{WardT2} for the non-invertible $\cT$ symmetry on $\amsmathbb{T}^2$ at $R^2=N$, and we show that it is satisfied.

To this end, we evaluate the partition function of a free scalar on $\amsmathbb{T}^2$, with momentum defects $\eta$ inserted along the two cycles of the torus,
\begin{equation}
    Z_{u,v}= \frac{R}{
    \abs{\eta(\tau)}^2
    }\sum_{m,w\in\integers}   e^{-\frac{\pi R^2}{\tau_2} \abs{(m+\frac{u}{2})-\tau (w+\frac{v}{2})}^2}~.
\end{equation}
To evaluate the above sum, we use the Poisson resummation formula
\begin{equation}
    \sum_{m\in\integers} e^{-\pi a(m+c)^2} = \frac{1}{\sqrt{a}}\sum_{m\in\integers} e^{ - \frac{\pi}{a}m^2 + 2\pi i c m}~,
\end{equation}
to obtain
\begin{align}
   Z_{u,v} &=  \frac{1}{
    \abs{\eta(\tau)}^2
    }\sum_{m,w\in\integers} e^{\frac{2\pi i m u}{N}} q^{\half p_-^2} \Bar{q}^{\half p_+^2}~,
\end{align}
where $q = e^{2\pi i \tau}$ and
\begin{align}
   p_\pm &= \frac{1}{\sqrt{2}}\left(\frac{m}{R} \pm  R (w+ \frac{v}{N})\right)~.
\end{align}
Note that the series converges absolutely, as $\tau_2>0$ so that $\abs{q}<1$. Differentiating with respect to $R$, we obtain
\begin{align}
f_{u,v} & \equiv \frac{\partial Z_{u,v}}{\partial R} = 
\frac{4\pi \tau_2}{R\abs{\eta(\tau)}^2}\sum_{m,w\in\integers}  p_-p_+  e^{\frac{2\pi i m u}{N}} q^{\half p_-^2}\Bar{q}^{\half p_+^2}\no \\
&=\frac{2\pi \sqrt{N}\tau_2}{\abs{\eta(\tau)}^2}\sum_{m,w\in\integers}  \left[\frac{m}{N} - \left(w+\frac{v}{N}\right)\right] \left[\frac{m}{N} + \left(w+\frac{v}{N}\right)\right]  e^{\frac{2\pi i m u}{N}} q^{\half p_-^2}\Bar{q}^{\half p_+^2} \,,
\end{align}
where we plugged in the value $R^2=N$. The selection rule \eqref{WardT2} on the torus is satisfied if
\begin{equation} \label{WardFuv}
    f_{0,0}  = -\frac{1}{N} \sum_{u,v=0}^{N-1} f_{u,v}~.
\end{equation}
The left-hand side of \eqref{WardFuv} is just 
\begin{align}
    f_{0,0}&=\frac{2\pi \sqrt{N}\tau_2}{\abs{\eta(\tau)}^2}\sum_{m,w\in\integers}  \left(\frac{m}{N} - w\right) \left(\frac{m}{N} +w\right) q^{\frac{N}{4} (\frac{m}{N}-w)^2}\Bar{q}^{\frac{N}{4} (\frac{m}{N}+w)^2}.
\end{align}
On right-hand side of \eqref{WardFuv}, the sum over $u$ gives 
\begin{align}
    \sum_{u=0}^{N-1} e^{\frac{2\pi i m u}{N}} = N\delta_{m = 0 \text{ mod }N},
\end{align}
so that the right-hand side reads
\begin{align}
    -\frac{1}{N} \sum_{u,v=0}^{N-1} f_{u,v} &= -\frac{2\pi \sqrt{N}\tau_2}{\abs{\eta(\tau)}^2}\sum_{v=0}^{N-1} \sum_{m,w\in\integers} \left(m - w-\frac{v}{N}\right) \left(m + w+\frac{v}{N}\right) q^{\frac{N}{4} (m+w+\frac{u}{N})^2}\Bar{q}^{\frac{N}{4} (m-w-\frac{u}{N})^2}\no\\
    &= -\frac{2\pi \sqrt{N}\tau_2}{\abs{\eta(\tau)}^2}\sum_{m',w'\in\integers} \left(m' + \frac{w'}{N}\right) \left(m' - \frac{w'}{N}\right) q^{\frac{N}{4} (m'+\frac{w'}{N})^2}\Bar{q}^{\frac{N}{4} (m'-\frac{w'}{N})^2}~,
\end{align}
where in the last step we defined $m' = N m$ and $w' = Nw+v$ with $v = 0, \dots, N-1$. Relabeling the indices as $m' = w$ and $w'=m$, we see that the selection rule \eqref{WardFuv} is satisfied. We have thus shown that the non-vanishing of the one-point function of $V_\s$ on $\amsmathbb{T}^2$ is compatible with the selection rule for the non-invertible $\cT$ symmetry. 

\newpage
\bibliographystyle{JHEP}
\bibliography{refs}
\end{document}